\documentclass[10pt,journal]{IEEEtran}

\ifCLASSINFOpdf

\else

\fi

\usepackage{color}
\usepackage{textcomp}
\usepackage[ruled]{algorithm2e}
\usepackage{graphicx}
\usepackage{subfigure}
\usepackage{amsmath,amssymb,amsfonts,mathrsfs,epsfig}
\usepackage{multirow}
\usepackage{amssymb}
\usepackage{algorithmic}
\usepackage{multirow}
\usepackage{changebar}
\usepackage{booktabs}
\usepackage{stfloats}
\usepackage{epstopdf}
\usepackage{xcolor,stfloats}
\usepackage{bm}

\usepackage{subeqnarray}
\usepackage{cases}

\newtheorem{remark}{Remark}

\begin{document}

\title{Introducing Meta-Fiber into Stacked Intelligent Metasurfaces for MIMO Communications: A Low-Complexity Design with only Two Layers}

\author{Hong Niu,~\IEEEmembership{}
        Jiancheng An,~\IEEEmembership{}
        Tuo Wu,~\IEEEmembership{}
        Jiangong Chen,~\IEEEmembership{}
        Yufei Zhao,~\IEEEmembership{}
        \\Yong Liang Guan,~\IEEEmembership{}
        Marco Di Renzo,~\IEEEmembership{Fellow, IEEE,}
        M\'{e}rouane Debbah,~\IEEEmembership{Fellow, IEEE,}
        \\George K. Karagiannidis,~\IEEEmembership{Fellow, IEEE,}
        H. Vincent Poor,~\IEEEmembership{Life Fellow, IEEE,}
        and Chau Yuen,~\IEEEmembership{Fellow, IEEE}

\thanks{
This work has been accepted for publication in IEEE Transactions on Wireless Communications, 2025.

H. Niu, J. An, T. Wu, Y. Zhao, Yongliang Guan, and C. Yuen are with the School of Electrical and Electronics Engineering, Nanyang Technological University, Singapore 639798 (email: hong.niu@ntu.edu.sg; jiancheng.an@ntu.edu.sg; tuo.wu@ntu.edu.sg; yufei.zhao@ntu.edu.sg; eylguan@ntu.edu.sg; chau.yuen@ntu.edu.sg).

J. Chen is with the Department of Electronic and Electrical Engineering, University College London, London, U.K. and he is also with with the National Key Laboratory of Wireless Communications, University of Electronic Science and Technology of China, Chengdu, China (email: jg\_chen@std.uestc.edu.cn).

M. Di Renzo is with Universit\'e Paris-Saclay, CNRS, CentraleSup\'elec, Laboratoire des Signaux et Syst\`emes, 3 Rue Joliot-Curie, 91192 Gif-sur-Yvette, France. (marco.di-renzo@universite-paris-saclay.fr), and with King's College London, Centre for Telecommunications Research -- Department of Engineering, WC2R 2LS London, United Kingdom (marco.di\_renzo@kcl.ac.uk).

M. Debbah is with the Center for 6G Technology, Khalifa University of Science and Technology, Abu Dhabi, United Arab Emirates (email: merouane.debbah@ku.ac.ae).

G. K. Karagiannidis is with the Department of Electrical and Computer Engineering, Aristotle University of Thessaloniki, 54124 Thessaloniki, Greece (email: geokarag@auth.gr).

H. V. Poor is with the Department of Electrical and Computer Engineering, Princeton University, Princeton, NJ, USA 08544 (email: poor@princeton.edu).

}
}


\markboth{IEEE}%
{Shell \MakeLowercase{\textit{et al.}}: }
\maketitle

\begin{abstract}
Stacked intelligent metasurfaces (SIMs), which integrate multiple programmable metasurface layers, have recently emerged as a promising technology for advanced wave-domain signal processing. SIMs benefit from flexible spatial degree-of-freedom (DoF) while reducing the requirement for costly radio-frequency (RF) chains. However, current state-of-the-art SIM designs face challenges such as complex phase shift optimization and energy attenuation from multiple layers. To address these aspects, we propose incorporating meta-fibers into SIMs, with the aim of reducing the number of layers and enhancing the energy efficiency. First, we introduce a meta-fiber-connected 2-layer SIM that exhibits the same flexible signal processing capabilities as conventional multi-layer structures, and we explains its operating principle. Subsequently, we formulate and solve the optimization problem of minimizing the mean square error (MSE) between the SIM channel and the desired channel matrices. Specifically, by designing the phase shifts of the meta-atoms associated with the transmitting-SIM and receiving-SIM, a non-interference system with parallel subchannels is established. In order to reduce the computational complexity, a closed-form expression for each phase shift at each iteration of an alternating optimization (AO) algorithm is proposed. We show that the proposed algorithm is applicable to conventional multi-layer SIMs. The channel capacity bound and computational complexity are analyzed to provide design insights. Finally, numerical results are illustrated, demonstrating that the proposed two-layer SIM with meta-fiber achieves over a 25$\%$ improvement in channel capacity while reducing the total number of meta-atoms by 59$\%$ as compared with a conventional seven-layer SIM.
\end{abstract}


\IEEEpeerreviewmaketitle

\vspace*{-3mm}
\section{Introduction}

\IEEEPARstart{R}{econfigurable} intelligent surfaces (RISs) \cite{RIS1}-\cite{b1} have recently emerged as a promising technology for sixth-generation (6G) wireless networks due to their ability to create high-quality channel environments \cite{b2}-\cite{b8}. By deploying low-cost nearly passive reconfigurable elements across a programmable metasurface, RISs can dynamically manipulate electromagnetic (EM) waves to reshape wireless channels proactively. Therefore, RISs enhance the transmission performance of wireless networks through the customization of adaptable propagation environments \cite{b10}-\cite{b18}.

With the deployment of a single-layer RIS, \emph{Huang et al.} \cite{Huang1} substantially improved the energy efficiency of the downlink in multiuser communication systems compared to conventional relaying solutions. Moreover, \emph{An et al.} \cite{pathloss1} designed a codebook-based framework to acquire accurate channel state information (CSI), making possible the deployment of numerous reconfigurable elements. However, existing research on RIS-assisted transmissions relies on a single-layer metasurface structure, which may not provide sufficient spatial degrees-of-freedom (DoFs) for complex signal processing operations \cite{Bai1}-\cite{b23}.

Recently, \emph{stacked intelligent metasurfaces (SIMs)} \cite{SIMo1}-\cite{SIM6} have been proposed as a promising technology for advanced wave-domain signal processing \cite{SIM7}-\cite{SIM8}. Specifically, \emph{Liu et al.} initially introduced a programmable diffractive deep neural network architecture based on multi-layer metasurfaces \cite{SIM3} to perform various signal processing tasks, such as image classification. Building on this concept, \emph{An et al.} \cite{SIM1} integrated SIMs with transceivers to perform signal processing tasks, such as precoding and combining. Due to the cascaded multi-layer structure, SIMs offer enhanced spatial DoFs while reducing the requirement for costly radio-frequency (RF) chains.

While the potential of SIM to enhance wireless communication performance is acknowledged in the literature, two significant challenges arise from its multi-layer structure. First, the multiple layers lead to an increased number of meta-atoms, resulting in numerous optimization variables that complicate system optimization. Second, as the signal propagates through each layer, it experiences inevitable energy attenuation \cite{Att1}, reducing the energy efficiency of transmission.


\begin{figure}
	\centering
	\includegraphics[width=3.5in]{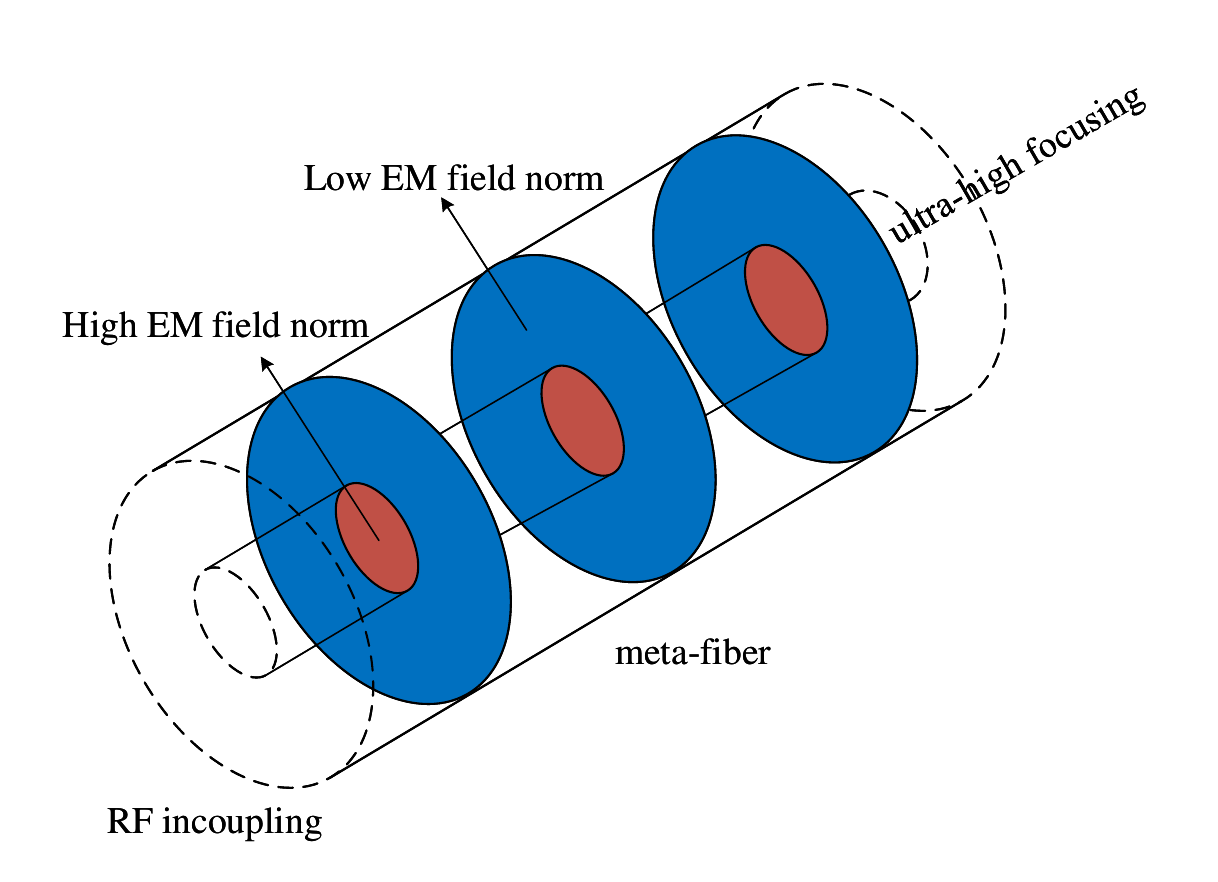}
	\vspace{-3mm}
	\caption{The concept of meta-fiber.}
    \label{fig_mfiber}
\end{figure}


In order to mitigate the attenuation caused by the multi-layer structure while retaining the advantages of optimization flexibility, it is beneficial to design the transmission topology within the SIM to minimize the number of layers. As illustrated in Fig. \ref{fig_mfiber}, the concept of meta-fiber is extended from optical fibers \cite{fiber1}-\cite{fiber2}, with the added capability of guiding RF signals. {The concentration of high EM field norm at the meta-fiber center provides enhanced energy-focusing capabilities, which enhances the coupling between connected atom pairs while suppressing unwanted leakage in the low EM field norm region, hence making interference from unconnected pairs negligible.} Integrating meta-fibers into the SIM structure enables the focusing of EM waves between fixed meta-atoms, thereby reducing the number of layers while still maintaining versatile signal processing capabilities in the wave domain. Specifically, \emph{Zhao et al.} \cite{fiber3} presented and summarized optical fiber-integrated metasurfaces with the current state of the art. \emph{Yang et al.} \cite{fiber4} discussed a platform and applications for optical fiber-integrated metasurfaces. Additionally, \emph{Sanjari et al.} \cite{fiber5} demonstrated an electronic-photonic metasurface that facilitates the conversion of an incident optical wave at 193 THz to a millimeter-wave signal at 28 GHz. In conclusion, the integration of metasurfaces and meta-fibers may be realized by leveraging fiber-integrated metasurface platforms and conversion techniques between optical and RF signals.

\begin{figure*}
	\centering
	\includegraphics[width=0.9\textwidth]{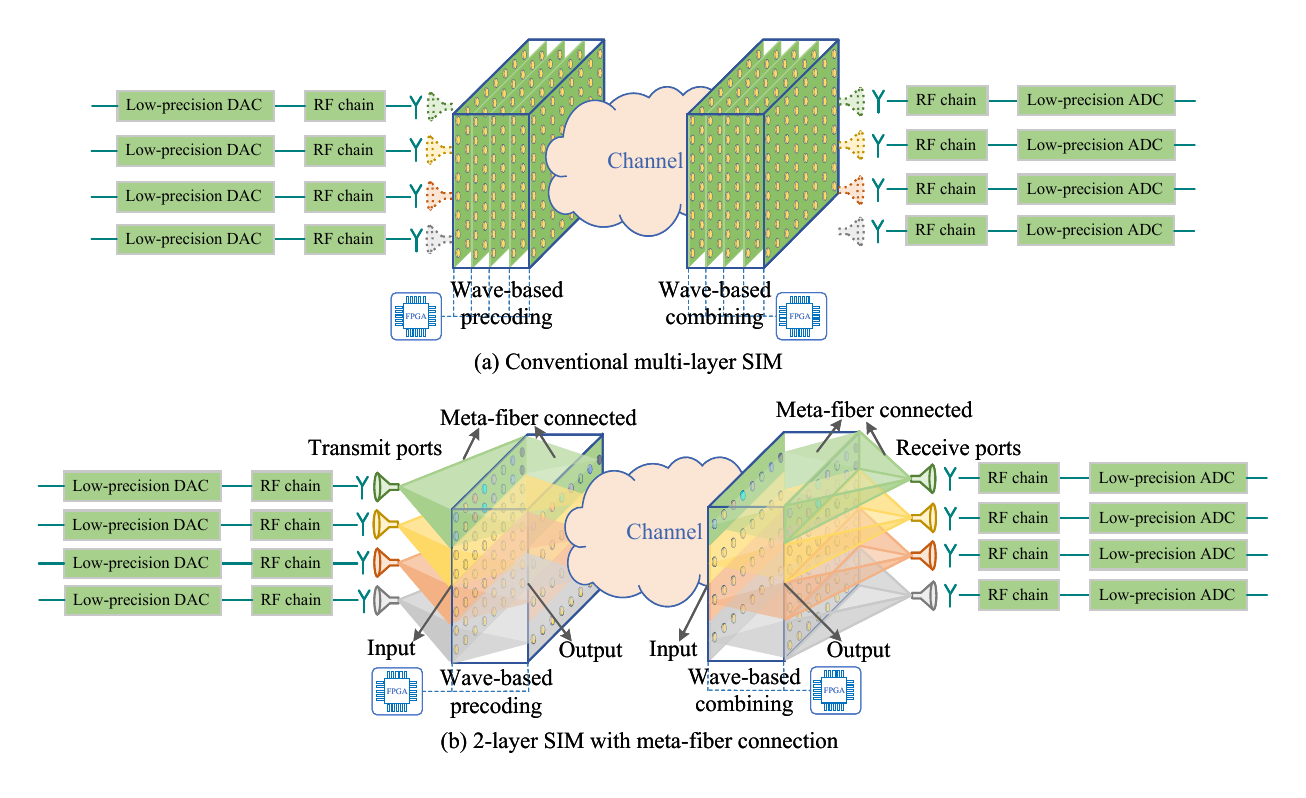}
	\vspace{-3mm}
	\caption{Comparison between the conventional multi-layer SIM and the proposed meta-fiber-connected 2-layer SIM for MIMO transmission.}
    \label{fig_1}
\end{figure*}

Motivated by the aforementioned considerations, we propose incorporating meta-fibers into the SIM structure, thereby reducing the number of layers, simplifying the design, and enhancing energy efficiency. As depicted in Fig. \ref{fig_1}, by introducing meta-fibers between the transmit antennas and the input layer of the SIM, as well as between the input and output layers of the SIM, a 2-layer SIM is proposed at the transmitter to maintain the same flexible signal processing capabilities as conventional multi-layer structures \cite{SIM1}. A similar design can be utilized at the receiver. Specifically, our main contributions are summarized as follows:
\begin{enumerate}

  \item \textbf{Novel topology:} We design the connection topology for the proposed 2-layer SIM as illustrated in Fig. \ref{fig_2}. Specifically, the SIM layer connected to the transmit and receive ports has a fully connected structure, while the internal structure of the SIM utilizes 2-to-1 partial connections. Moreover, the operating principle of the proposed 2-layer SIM is illustrated to explain why it maintains the same flexible signal processing capabilities as conventional multi-layer structures.

  \item \textbf{Closed-form solution:} We propose an algorithm to optimize the phase shifts of the transmitting and receiving SIMs. In order to reduce the computational complexity, we derive efficient closed-form expressions for the phase shifts at each iteration of the algorithm, which is shown to converge to a Karush-Kuhn-Tucker (KKT) solution of the non-convex problem. Furthermore, we extend the derived closed-form solution to the conventional multi-layer SIM architecture, demonstrating the generality and versatility of the derived solution.

  \item \textbf{Theoretical analysis:} We analyze the channel capacity bound and computational complexity. The upper bound of the channel capacity provides insights on the working principle of the proposed meta-fiber SIM and how multiple parallel single-input single-output (SISO) sub-channels are realized between the paired ports. The computational complexity analysis shows that the derived closed-form solutions exhibit linear complexity as a function of the number of meta-atoms per layer.

  \item \textbf{Performance evaluation:} We illustrate simulation results to demonstrate the superiority of the proposed 2-layer SIM and the accuracy of the obtained analytical upper bounds for the channel capacity. Additionally, we verify the substantial performance improvements of the proposed meta-fiber connected 2-layer SIM compared to a conventional 7-layer SIM.

\end{enumerate}

The rest of this paper is organized as follows. In Section II, we present the connection topology and fundamental operating principles of the meta-fiber-connected 2-layer SIM. In Section III, we formulate the SIM-aided channel optimization problem and derive efficient closed-form expressions for the phase shifts. Moreover, we extend the derived closed-form expressions to the conventional multi-layer SIM, demonstrating the versatility of the solution proposed in Section IV. Additionally, we analyze the channel capacity bound and computational complexity in Section V. Numerical results are provided in Section VI. Finally, concluding remarks are presented in Section VII.

\begin{table}[h]
\center
\caption{Notation List}
\vspace{+5mm}
\begin{tabular}{cc}
\toprule
Symbol&Meaning\\
  \midrule
  ${{\mathbf{I}}_S}$&$S\times S$ identity matrix\\
  $j$&Imaginary unit of a complex number\\
  ${\left(  \cdot  \right)^H}$&Conjugate transpose\\
  $\left| x \right|$&Modulus of a complex number $x$\\
  $\left\| {\mathbf{x}} \right\|$&2-norm of a vector ${\mathbf{x}}$\\
  $\left\| {\mathbf{X}} \right\|_F$&Frobenius norm of a matrix ${\mathbf{X}}$\\
  ${\text{diag}}\left(  \cdot  \right)$&Diagonal matrix\\
  ${{\mathbf{1}}_N}$&$N$-dimensional all-one vector\\
  ${x_i}$&$i$-th element of vector ${\mathbf{x}}$\\
  ${X_{i,j}}$&$(i,j)$ element of matrix $\mathbf X$\\
  ${X_{i,j,k}}$&$(i,j,k)$ element of tensor $\mathbf X$\\
  $\sum\limits_{i = 1}^N {{x_i}} $&The sum of $x_i$ from $i=1$ to $N$\\
  $\partial f/\partial x$&Partial derivative of function $f$ with respect to $x$\\
  ${\log _a}\left(  \cdot  \right)$&Logarithmic function with base $a$\\
  $\cos $&Cosine function\\
  $\arccos $&Inverse cosine function\\
  $\arctan $&Inverse tangent function\\
  $\mathbb{C}$&Complex number field\\
  ${\mathbb{R}^ + }$&Positive real number field\\
  $\mathcal{O}$&Magnitude of computational complexity\\
  $ \otimes $&Kronecker product\\
  ${\text{i}}{\text{.i}}{\text{.d}}{\text{.}}$&Independent identically distributed\\
  \multirow{2}{*}{$ \sim \mathcal{C}\mathcal{N}\left( {{\mathbf{\mu }},{\mathbf{\Sigma }}} \right)$}&{Complex Gaussian distribution}\\&{ with mean vector ${\mathbf{\mu }}$ and covariance matrix ${\mathbf{\Sigma }}$}\\
  \bottomrule
\end{tabular}
\label{tbl}
\end{table}

\textit{Notation}: In the following, bold lower-case and upper-case letters denote vectors and matrices, respectively. The notation utilized is summarized in Table I.

\section{System Model}

In this section, we elaborate on the fundamental principles of meta-fiber-connected 2-layer SIMs. Furthermore, we explain why a 2-layer SIM based on meta-fibers benefit from signal processing capabilities similar to those of the conventional multi-layer architecture.

\subsection{Meta-Fiber-Connected 2-layer SIM}

First of all, we review the conventional multi-layer SIM structure depicted in Fig. \ref{fig_1}(a). As shown in \cite{SIM1}, with the deployment of several metasurface layers, an SIM is capable of performing advanced processing tasks, e.g., channel diagonalization. Specifically, an SIM is integrated at both the transmitter (TX) and receiver (RX) sides to perform wave-domain precoding and combining, respectively. Both the TX-SIM and RX-SIM are composed of multiple stacked metasurface layers and each layer consists of a given number of meta-atoms. By adjusting the drive level of the control circuit associated with each meta-atom, the phase shift of the EM wave signal passing through each meta-atom can be precisely controlled. Furthermore, all meta-atoms are connected to a field programmable gate array (FPGA) board, through which the joint control of all the atoms enables the SIM to generate customized spatial waveforms at the output layer, i.e., to perform signal processing in the wave domain.

The SIM architecture is attractive due to the reduced number of RF chains and low-precision digital-to-analog converters (DACs)/ analog-to-digital converters (ADCs) that are required at TX and RX. However, conventional multi-layer SIM structures suffer from optimization complexity due to the large number of total meta-atoms. Furthermore, in practical implementations, EM waves generally suffer 10$\%$ or more energy attenuation when passing through each layer of the SIM, primarily due to fabrication challenges and material losses \cite{fiber3}, which limit the number of stacked metasurface layers.

Motivated by these considerations, we aim to introduce a new SIM architecture with a limited number of layers. Specifically, we focus on a 2-layer SIM structure, as this offers two key benefits. On the one hand, reducing the number of layers decreases the total number of meta-atoms, thereby reducing the computational complexity of the SIM design. On the other hand, this also decreases energy attenuation. {Specifically, assuming 10$\%$ energy attenuation per layer, the energy attenuation caused by a 7-layer SIM is approximately 52.2$\%$ ($1 - {\left( {1 - 0.1} \right)^7}$), far exceeding the 19$\%$ ($1 - {\left( {1 - 0.1} \right)^2}$) of a 2-layer SIM. Furthermore, a 2-layer structure is the optimal configuration that is capable of achieving wave-domain signal processing operations, as a 1-layer architecture only modifies the phase, not the amplitude, of the EM signal. Although multi-layer structures with more than three layers are also theoretically feasible, this paper focus on 2-layer topologies to reduce energy attenuation.}

To ensure that the proposed 2-layer SIM has the necessary signal processing capabilities in the wave domain \cite{SIMo1}, meta-fibers are introduced to control the transmission coefficients between connected meta-atoms. The use of meta-fibers is beneficial for two main reasons:

1) The diffraction coefficients between meta-atoms in multi-layer SIM are determined by Rayleigh-Sommerfeld's diffraction theory, which may be difficult to control. By contrast, the introduction of meta-fibers facilitate the control of diffraction coefficients in terms of amplitude and phase variation.

2) The meta-atoms connected via the meta-fibers are minimally affected by interference generated by other meta-atoms, enhancing the performance of the SIM.

\begin{figure}[t]
\centering
\includegraphics[width=3.5in]{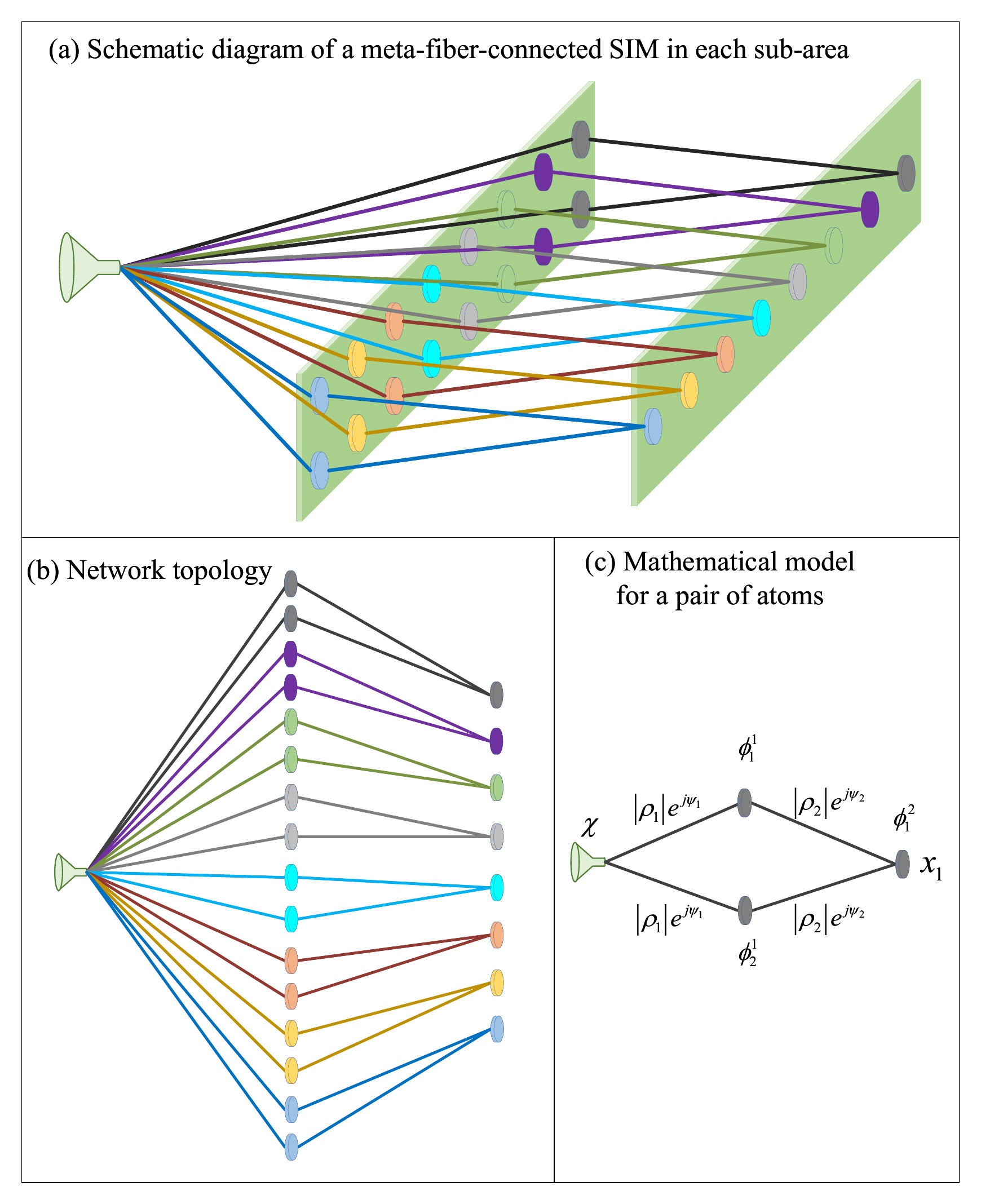}
\vspace{-3mm}
\caption{Network topology and mathematical model for each sub-area.}
\label{fig_2}
\vspace{-0em}
\end{figure}

In view of these considerations, we proposed a meta-fiber-connected 2-layer SIM for multiple-input multiple-output (MIMO) transmission, as shown in Fig. \ref{fig_1}(b). Similar to the conventional multi-layer SIM, the proposed meta-fiber-connected 2-layer SIM can perform advanced precoding and combining operations in the wave domain, but at lower computational complexity and reduced energy attenuation.

As observed from Fig. \ref{fig_1}(b), we divide the two layers of the SIM into several sub-areas, such that the meta-atoms in each sub-area and the corresponding transmission port are connected via meta-fibers. In order to simplify the structure and avoid interference, different sub-areas are assumed to be unconnected, and the transmission ports are only connected to the input layer of TX-SIM.

Explicitly, a schematic diagram of the meta-fiber-connected SIM in each sub-area is depicted in Fig. \ref{fig_2}(a). As depicted in the figure, the network topology in each sub-area is the same as a single-layer perceptron architecture in typical neural network models \cite{Per1,Per2}. Specifically, each sub-area is comprised of a transmission port and two layers of meta-atoms. In terms of network topology, each transmission port is fully connected to the input layer, while every two meta-atoms in the input layer are connected to one meta-atom in the output layer. As a result, the number of meta-atoms in the input layer is twice that of the output layer. The mathematical model is illustrated in Fig. \ref{fig_2}(c). In terms of wave domain processing, the objective is to synthesize an output symbol ${x_1}$ with any magnitude and phase. As illustrated in Fig. \ref{fig_2}(c), the output symbol can be expressed as
\begin{equation}
{x_1} = \chi  \left| {{\rho _1}} \right|\left| {{\rho _2}} \right|{e^{j\left( {{\psi _1} + {\psi _2}} \right)}}\left( {\phi _1^1 + \phi _2^1} \right)\phi _1^2,
\end{equation}
where $\chi  \in {\mathbb{R}^ + }$ denotes the amplitude of the input signal at the transmission port, $\left| {{\rho _1}} \right|$ and $\left| {{\rho _2}} \right|$ represent the amplitudes of the channel coefficients introduced by the meta-fiber from the transmission port to the input layer and from the input layer to the output layer, respectively, while ${{\psi _1}}$ and ${{\psi _2}}$ are the corresponding phase variations\footnote{In practical scenarios, the meta-fibers introduce signal amplitude attenuations and phase variations, typically influenced by factors such as material properties and length. To facilitate the analysis, we consider that two different types of meta-fibers are utilized for the first and second layers of the SIM. {In general, the number of distinguishable meta-fiber types is limited by fabrication cost. Although this paper considers two types, a similar functionality can be achieved with a single type, as shown in the simulation results.}}. Moreover, ${\phi _1^1}$ and ${\phi _2^1}$ stand for the transmission coefficients imposed by the two meta-atoms on the input layer of the SIM, while ${\phi _1^2}$ represents the transmission coefficient imposed by the meta-atom on the output layer.
In this paper, we assume that the meta-atoms are able to alter the phase of the input signals, i.e., $\phi _m^l = {e^{j\theta _m^l}}$, where ${\phi _m^l}$ denotes the transmission coefficient imposed by the $m$-th meta-atoms on the $l$-th layer with $\theta _m^l \in \left[ {0,2\pi } \right)$ representing the phase shift.

It is desirable that ${x_1}$ has an arbitrary adjustable amplitude and phase, i.e., ${x_1} = \left| {{x_1}} \right|{e^{j{\delta _1}}}$, where $\left| {{x_1}} \right| \in \left[ {0,A} \right]$ stands for any adjustable magnitude, while ${\delta _1} \in \left[ {0,2\pi } \right)$ denotes any adjustable phase. To this end and to facilitate the design, the meta-atoms in the input layer are designed to optimize the amplitude, while those in the output layer are designed to optimize the phase shift. This design criterion can be formulated as follows:
\begin{subequations}\label{layer:1}
\begin{align}
 & \theta _1^1 = \arccos \left( {\frac{b}{2}} \right), \label{layer1:1}\\
 & \theta _2^1 =  - \arccos \left( {\frac{b}{2}} \right), \label{layer1:2}\\
 & \theta _1^2 = {\delta _1} - {\psi _1} - {\psi _2}, \label{layer2:1}
\end{align}
\end{subequations}
where $b = \left| {{x_1}} \right|/\left( {\chi \left| {{\rho _1}} \right|\left| {{\rho _2}} \right|} \right)$.

Through the design in (\ref{layer:1}), flexible waveform customization can be obtained. Since $\left| {{x_1}} \right|$ varies in the range $\left[ {0,A} \right]$, $b$ varies in the range of $\left[ {0,A/\left( {\chi \left| {{\rho _1}} \right|\left| {{\rho _2}} \right|} \right)} \right]$. By definition, the domain of the inverse cosine function in (\ref{layer1:1}) and (\ref{layer1:2}) imposes the constraint $\frac{b}{2} \in \left[ {0,1} \right]$, so that the largest amplitude of $x_1$ is
\begin{equation}\label{constraint:1}
A = 2\chi \left| {{\rho _1}} \right|\left| {{\rho _2}} \right|.
\end{equation}
Eq. (\ref{constraint:1}) implies that the maximum amplitude of the output symbol is proportional to the amplitude of the input signal at the transmission port.

Considering Fig. \ref{fig_1}(b) as reference, the number of transmission ports is denoted by $S$, and the set of all transmission ports is represented by $\mathcal{S} = \left\{ {1,2, \cdots ,S} \right\}$. As far as the connection topology is concerned, each transmission port in a sub-area is connected to multiple meta-atoms via meta-fibers. The number of meta-atoms on the input and output layers of TX-SIM is denoted by $2M$ and $M$, respectively. Considering the network topology in Fig. \ref{fig_2}, the transmission coefficient matrix between the transmission ports and the input layer of TX-SIM can be expressed as
\begin{equation}\label{TCM:1}
{{\mathbf{W}}^1} = \left| {{\rho _1}} \right|{e^{j{\psi _1}}}{\text{diag}}(\underbrace {{{\mathbf{1}}_{2M}},{{\mathbf{1}}_{2M}}, \cdots ,{{\mathbf{1}}_{2M}}}_S) \in {\mathbb{C}^{2MS \times S}},
\end{equation}
where ${{\mathbf{1}}_{2M}}$ denotes the $2M$-dimensional all-ones vector. Eq. (4) formulates a fully connected structure between the transmission port and the input layer of TX-SIM in each sub-area, and the decoupling between the sub-areas imposed by the block-diagonal structure. Similarly, the transmission coefficient matrix between the input layer and the output layer of TX-SIM can be formulated as
\begin{equation}\label{TCM:2}
{{\mathbf{W}}^2} = \left| {{\rho _2}} \right|{e^{j{\psi _2}}}{\text{diag}}(\underbrace {{\mathbf{1}}_2^H,{\mathbf{1}}_2^H, \cdots ,{\mathbf{1}}_2^H}_{MS}) \in {\mathbb{C}^{MS \times 2MS}},
\end{equation}
where ${{\mathbf{1}}_2^H}$ represents the $2$-dimensional all-ones row vector, indicating that every pair of meta-atoms in the input layer is connected to one meta-atom in the output layer via the meta-fibers. Therefore, the end-to-end precoding matrix of TX-SIM is given by
\begin{equation}\label{TCM:3}
{\mathbf{P}} = {{\mathbf{\Phi }}^2}{{\mathbf{W}}^2}{{\mathbf{\Phi }}^1}{{\mathbf{W}}^1} \in {\mathbb{C}^{MS \times S}},
\end{equation}
where ${{\mathbf{\Phi }}^1} = {\text{diag}}\left( {\phi _1^1,\phi _2^1, \cdots ,\phi _{2MS}^1} \right) \in {\mathbb{C}^{2MS \times 2MS}}$ stands for the transmission coefficient matrix of the meta-atoms on the input layer, and ${{\mathbf{\Phi }}^2} = {\text{diag}}\left( {\phi _1^2,\phi _2^1, \cdots ,\phi _{MS}^2} \right) \in {\mathbb{C}^{MS \times MS}}$ is the transmission coefficient matrix of the meta-atoms on the output layer.

A similar end-to-end decoding matrix can be derived for the RX-SIM. Specifically, denoting the number of meta-atoms on the input layer by $N$, and the number of meta-atoms on the output layer by $2N$, the transmission coefficient matrix between the input layer and the output layer of RX-SIM can be expressed as
\begin{equation}\label{RCM:1}
{{\mathbf{U}}^2} = \left| {{\rho _2}} \right|{e^{j{\psi _2}}}{\text{diag}}(\underbrace {{{\mathbf{1}}_2},{{\mathbf{1}}_2}, \cdots ,{{\mathbf{1}}_2}}_{NS}) \in {\mathbb{C}^{2NS \times NS}}.
\end{equation}
Furthermore, the transmission coefficient matrix between the output layer of RX-SIM and the receive ports can be formulated as
\begin{equation}\label{RCM:2}
{{\mathbf{U}}^1} = \left| {{\rho _1}} \right|{e^{j{\psi _1}}}{\text{diag}}(\underbrace {{\mathbf{1}}_{2M}^H,{\mathbf{1}}_{2M}^H, \cdots ,{\mathbf{1}}_{2N}^H}_S) \in {\mathbb{C}^{S \times 2NS}}.
\end{equation}
Therefore, the end-to-end combining matrix of RX-SIM is given by
\begin{equation}\label{RCM:3}
{\mathbf{Q}} = {{\mathbf{U}}^1}{{\mathbf{\Psi }}^1}{{\mathbf{U}}^2}{{\mathbf{\Psi }}^2} \in {\mathbb{C}^{S \times NS}},
\end{equation}
where ${{\mathbf{\Psi }}^1} = {\text{diag}}\left( {\varphi _1^1,\varphi _2^1, \cdots ,\varphi _{2NS}^1} \right) \in {\mathbb{C}^{2NS \times 2NS}}$ denotes the transmission coefficient matrix of the meta-atoms on the output layer of RX-SIM, and ${{\mathbf{\Psi }}^2} = {\text{diag}}\left( {\varphi _1^2,\varphi _2^2, \cdots ,\varphi _{NS}^2} \right) \in {\mathbb{C}^{NS \times NS}}$ represents the transmission coefficient matrix imposed by the input layer. Similar to TX-SIM, we assume that the meta-atoms of RX-SIM are able to alter only the phase of the input signals, i.e., $\varphi _n^k = {e^{j\vartheta _n^k}}$, where ${\varphi _n^k}$ denotes the transmission coefficient imposed by the $n$-th meta-atom on the $k$-th layer of RX-SIM with $\vartheta _n^k \in \left[ {0,2\pi } \right)$ representing the phase shift.

Regarding the MIMO channel between the output layer of TX-SIM and the input layer of RX-SIM, we consider the canonical independent and identically distributed (i.i.d.) Rayleigh fading channel, i.e., \cite{R2,PL1}
\begin{equation}\label{Gchannel:1}
{\mathbf{G}} \sim \mathcal{C}\mathcal{N}\left( {{\mathbf{0}},\beta {{\mathbf{I}}_{2NS}} \otimes {{\mathbf{I}}_{MS}}} \right),
\end{equation}
where $\beta  = {\beta _0}{d^{ - \gamma }}$ denotes the path loss with ${\beta _0}$, $d$, and $\gamma $ being the free space path loss at the reference distance of 1 m, the distance between the output layer of TX-SIM and the input layer of RX-SIM, and the path loss exponent \cite{pathloss1}, respectively.

{Although the proposed algorithm is developed under a quasi-static Rayleigh fading assumption (i.e., the channel remains constant within each coherence time slot), it is inherently compatible with dynamic environments. Specifically, the algorithm can be re-executed at each channel update or adapted using statistical CSI \cite{Sta1}, making it suitable for scenarios with either CSI refresh or statistical CSI availability.}

\begin{remark}
By direct inspection of the signal model for the proposed meta-fiber connected SIM, we evince that it is quite different from the signal model of a conventional SIM. In the latter case, the propagation channels between pairs of meta-atoms are determined by Rayleigh-Sommerfield's diffraction, which depend on the positions of the meta-atoms and notably the inter-layer transmission distances. This diffraction model imposes that all the meta-atoms are connected to one another and introduces larger signal attenuations as the number of layers increases. In the proposed meta-fiber connected SIM architecture, on the other hand, the meta-atoms are connected through meta-fibers, which reduces the attenuation of signals and alleviates the impact of the exact placement of the meta-atoms. Also, the sub-areas of the SIM are decoupled, simplifying the optimization of the whole SIM.

{Eqs. (\ref{TCM:3}), (\ref{RCM:3}), and (\ref{Gchannel:1}) rely on three assumptions: 1) continuous phase coefficients, 2) high transmission efficiency of meta-fibers relative to wireless propagation, and 3) perfect CSI available at both TX-SIM and RX-SIM for phase shift design. These assumptions are theoretically sound and supported by practical observations. Notably, compared to conventional multi-layer SIM architectures, the proposed 2-layer SIM may retain performance advantages even under discrete phase constraints \cite{pathloss1} or in the presence of channel estimation errors \cite{CEE1,CEE2}.}
\end{remark}

\subsection{Channel Diagonalization}

In this subsection, through the precoding of TX-SIM and the combining of RX-SIM, channel diagonalization enables the transmission and recovery of multiple data streams through the corresponding transmit and receive ports. The detailed procedures are outlined below.

To begin with, the singular value decomposition (SVD) is applied to the MIMO channel ${\mathbf{G}}$ such that
\begin{equation}\label{SVD:1}
{\mathbf{G}} = {\mathbf{DS}}{{\mathbf{V}}^H},
\end{equation}
where ${\mathbf{D}} \in {\mathbb{C}^{2NS \times 2NS}}$ and ${\mathbf{V}} \in {\mathbb{C}^{MS \times MS}}$ are unitary matrix satisfying ${{\mathbf{D}}^H}{\mathbf{D}} = {{\mathbf{I}}_{2NS}}$ and ${{\mathbf{V}}^H}{\mathbf{V}} = {{\mathbf{I}}_{MS}}$, respectively. Moreover, ${\mathbf{S}} = \left[ {\begin{array}{*{20}{c}}
  {\mathbf{\Sigma }}&{\mathbf{0}} \\
  {\mathbf{0}}&{\mathbf{0}}
\end{array}} \right] \in {\mathbb{C}^{2NS \times MS}}$ represents the matrix of singular values with diagonal matrix ${\mathbf{\Sigma }} = {\text{diag}}\left( {{\lambda _1},{\lambda _2}, \cdots ,{\lambda _R}} \right) \in {\mathbb{C}^{R \times R}}$, whose entries arranged in non-increasing order ${\lambda _1} \geqslant {\lambda _2} \geqslant  \cdots  \geqslant {\lambda _R}$, where $R \leqslant \min \left( {MS,2NS} \right)$ is the rank of channel matrix ${\mathbf{G}}$.

Furthermore, by implementing the precoding and combining through the TX-SIM and RX-SIM, i.e.,
\begin{equation}
{\mathbf{P}} \to {{\mathbf{V}}_{:,1:S}}{\mathbf{S}}_{1:S,1:S}^{ - 1/2},
\end{equation}
\begin{equation}
{\mathbf{Q}} \to {\mathbf{S}}_{1:S,1:S}^{ - 1/2}{\mathbf{D}}_{:,1:S}^H,
\end{equation}
the SIM channel matrix degenerates into the identity matrix as
\begin{equation}\label{UM:1}
{\mathbf{QGP}} \to {{\mathbf{I}}_S}.
\end{equation}

\begin{remark}
As inferred from (\ref{UM:1}), multiple data streams are directly conveyed through the corresponding transmit and receive ports, significantly reducing the complexity for transmission and reception. Compared with the work in \cite{SIM1}, the proposed scheme further balances the transmission quality of each data stream, minimizing the probability of poor channel conditions.
\end{remark}

\section{Problem Formulation and Solution}

In this section, we formulate the problem of minimizing the mean square error (MSE) between the SIM channel and the desired identity matrix by optimizing the phase shifts of the TX-SIM and RX-SIM. Furthermore, we derived closed-form expressions for each phase shift while keeping the other phase shifts fixed. Finally, we propose an AO algorithm to obtain final solutions that satisfy the KKT condition of the formulated problem.

\subsection{Problem Formulation}

Based on (\ref{UM:1}), the optimization problem is formulated as

\begin{subequations}\label{pf:1}
\begin{align}
\mathop {\min }\limits_{\phi _m^l,\varphi _n^k,\alpha } &J = \left\| {{\mathbf{QGP}} - \alpha {{\mathbf{I}}_S}} \right\|_F^2, \label{Za}\\
{\text{s}}{\text{.t}}{\text{.}} \;\;\;\; &{\mathbf{P}} = {{\mathbf{\Phi }}^2}{{\mathbf{W}}^2}{{\mathbf{\Phi }}^1}{{\mathbf{W}}^1}, \label{Zb}\\
&{\mathbf{Q}} = {{\mathbf{U}}^1}{{\mathbf{\Psi }}^1}{{\mathbf{U}}^2}{{\mathbf{\Psi }}^2},\label{Zc}\\
&\left| {\phi _m^1} \right| = 1,m \in \left\{ {1,2, \cdots ,2MS} \right\},\label{Zd}\\
&\left| {\phi _m^2} \right| = 1,m \in \left\{ {1,2, \cdots ,MS} \right\},\label{Ze}\\
&\left| {\varphi _n^1} \right| = 1,n \in \left\{ {1,2, \cdots ,2NS} \right\},\label{Zf}\\
&\left| {\varphi _n^2} \right| = 1,n \in \left\{ {1,2, \cdots ,NS} \right\},\label{Zg}\\
&\alpha  \in {\mathbb{R}^*},\label{Zh}
\end{align}
\end{subequations}
where $\alpha $ is an equivalent channel gain, representing the quality of the equivalent channel.
The objective function in (\ref{Za}) aims to provide an identity channel matrix through the precoding at TX-SIM and the combining at RX-SIM, as described in (\ref{UM:1}). Moreover, the constraints in (\ref{Zb}) and (\ref{Zc}) quantify the end-to-end channel of TX-SIM and RX-SIM, as shown in (\ref{TCM:3}) and (\ref{RCM:3}). Furthermore, the constraints in (\ref{Zb}), (\ref{Ze}), (\ref{Zf}), and (\ref{Zg}) indicate that the influence of meta-atoms primarily affects the phase shift rather than the amplitude, with different numbers of meta-atoms in each layer due to the designed non-uniform network topology.

Due to the non-convex constant modulus constraint in (\ref{Zb}), (\ref{Ze}), (\ref{Zf}), and (\ref{Zg}), along with the highly coupled variables in (\ref{Za}), deriving optimal solutions for Problem (\ref{pf:1}) is non-trivial. Although some optimization algorithms can approximate the optimal solution, their accuracy and robustness cannot be guaranteed. Therefore, in this paper, we derive closed-form expressions for each variables by setting the partial derivative to zero. The derived results offers intuitive insight for the design of SIM and benefits from a lower computational complexity compared to conventional optimization algorithms.

\subsection{Closed-form Solution for Each Variable}

In this subsection, an efficient closed-form solution is derived for each variable by calculating its partial derivative and solving for the roots that set the partial derivatives to zero. The objective function in (\ref{Za}) with respect to the phase shifts of the $q$-th layer\footnote{Note that the superscript $q$ is labeled by treating TX-SIM and RX-SIM as a unified entity. Specifically, $q=1,2,3$, and 4 represent the input layer of TX-SIM, the output layer of TX-SIM, the input layer of RX-SIM, and the output layer of RX-SIM, respectively. Therefore, we have ${{\mathbf{\Phi }}^3} = {{\mathbf{\Psi }}^2},{{\mathbf{\Phi }}^4} = {{\mathbf{\Psi }}^1}$.} is rewritten as
\begin{equation}\label{Re:1}
J = \left\| {{{\mathbf{R}}^q}{{\mathbf{\Phi }}^q}{{\mathbf{T}}^q} - {\mathbf{I}}} \right\|_F^2,
\end{equation}
where ${{\mathbf{R}}^q},q \in \left\{ {1,2,3,4} \right\}$ denotes the cascaded channel spanning from the $q$-th layer to the receive ports, expressed as

\begin{equation}
{{\mathbf{R}}^1} = {\mathbf{QG}}{{\mathbf{\Phi }}^2}{{\mathbf{W}}^2}, {{\mathbf{R}}^2} = {\mathbf{QG}}, {{\mathbf{R}}^3} = {{\mathbf{U}}^1}{{\mathbf{\Psi }}^1}{{\mathbf{U}}^2}, {{\mathbf{R}}^4} = {{\mathbf{U}}^1},
\end{equation}
${{\mathbf{T}}^q},q \in \left\{ {1,2,3,4} \right\}$ represents the cascaded channel spanning from the transmit ports to the $q$-th layer (excluding the meta-atoms), formulated by

\begin{equation}
{{\mathbf{T}}^1} = {{\mathbf{W}}^1}, {{\mathbf{T}}^2} = {{\mathbf{W}}^2}{{\mathbf{\Phi }}^1}{{\mathbf{W}}^1}, {{\mathbf{T}}^3} = {\mathbf{GP}}, {{\mathbf{T}}^4} = {{\mathbf{U}}^2}{{\mathbf{\Psi }}^2}{\mathbf{GP}},
\end{equation}
${\mathbf{I}} = \alpha {{\mathbf{I}}_S}$ indicates the desired fitting channel matrix, and ${{\mathbf{\Phi }}^q} = {\text{diag}}\left( {{{\left[ {{e^{j\theta _1^q}},{e^{j\theta _2^q}}, \cdots ,{e^{j\theta _{{Q_q}}^q}}} \right]}^T}} \right) \in {\mathbb{C}^{{Q_q} \times {Q_q}}}$ stands for the transmission coefficient matrix imposed by the $q$-th layer with ${Q_q}$ denoting the total number of meta-atoms thereon, given ${Q_1} = 2MS,{Q_2} = MS,{Q_3} = NS$, and ${Q_4} = 2NS$.

By definition, the objective function of (\ref{Re:1}) can be expressed as
\begin{equation}\label{Re:2}
J = \sum\limits_{k = 1}^S {\sum\limits_{j = 1}^S {{{\left| {\sum\limits_{i = 1}^{{Q_q}} {R_{k,i}^qT_{i,j}^q{e^{j\theta _i^q}}}  - {I_{k,j}}} \right|}^2}} },
\end{equation}
where ${R_{k,i}^q}$ denotes the element in the $k$-th row and $j$-th column of the matrix ${{\mathbf{R}}^q}$, and $T_{i,j}^q$ represents the element in the $i$-th row and $j$-th column of the matrix ${{\mathbf{T}}^q}$.

In order to simplify the expression in (\ref{Re:2}), tensors ${{\mathbf{L}}^q} \in {\mathbb{C}^{S \times {Q_q} \times S}}$ are defined as
\begin{equation}
L_{k,i,j}^q = R_{k,i}^qT_{i,j}^q.
\end{equation}

As such, the expression in (\ref{Re:2}) is simplified as
\begin{equation}\label{Re:3}
J = \sum\limits_{k = 1}^S {\sum\limits_{j = 1}^S {{{\left| {\sum\limits_{i = 1}^{{Q_q}} {\left| {L_{k,i,j}^q} \right|{e^{j\left( {\xi _{k,i,j}^q + \theta _i^q} \right)}}}  - \left| {{I_{k,j}}} \right|{e^{j{\zeta _{k,j}}}}} \right|}^2}} } ,
\end{equation}
where ${\left| {L_{k,i,j}^q} \right|}$ and ${\xi _{k,i,j}^q}$ denote the amplitude and phase of ${L_{k,i,j}^q}$, i.e., $L_{k,i,j}^q = \left| {L_{k,i,j}^q} \right|{e^{j\xi _{k,i,j}^q}}$, while ${\left| {{I_{k,j}}} \right|}$ and ${{\zeta _{k,j}}}$ denote those of ${{I_{k,j}}}$, i.e., ${{I_{k,j}} = \left| {{I_{k,j}}} \right|{e^{j{\zeta _{k,j}}}}}$.

Furthermore, (\ref{Re:3}) can be unfolded as
\begin{equation}\label{Re:4}
\begin{gathered}
  J = \sum\limits_{k = 1}^S {\sum\limits_{j = 1}^S {\left( {{{\left| {{I_{k,j}}} \right|}^2} + \sum\limits_{i = 1}^{{Q_q}} {{{\left| {L_{k,i,j}^q} \right|}^2}} } \right)} }  \hfill \\
   + 2\sum\limits_{k = 1}^S {\sum\limits_{j = 1}^S {\sum\limits_{l = i + 1}^{{Q_q}} {\sum\limits_{i = 1}^{{Q_q} - 1} {\left| {L_{k,i,j}^q} \right|\left| {L_{k,l,j}^q} \right|} } } }  \hfill \\
  \cos \left( {\xi _{k,i,j}^q + \theta _i^q - \xi _{k,l,j}^q - \theta _l^q} \right) \hfill \\
   - 2\sum\limits_{k = 1}^S {\sum\limits_{j = 1}^S {\sum\limits_{i = 1}^{{Q_q}} {\left| {L_{k,i,j}^q} \right|\left| {{I_{k,j}}} \right|} \cos \left( {\xi _{k,i,j}^q + \theta _i^q - {\zeta _{k,j}}} \right)} } . \hfill \\
\end{gathered}
\end{equation}

{To derive the optimal solution for the $m$-th phase shift of the $q$-th layer $\theta _m^q$, we compute the partial derivative of the objective function $J$ in (22) with respect to $\theta _m^q$. Notably, only the second and third terms in (22) depend on $\theta _m^q$. In particular, the second term requires a case-by-case analysis depending on whether the subscripts $l$ and $i$ are equal to $m$.} Therefore, the partial derivative of $J$ with respect to $\theta _m^q$ is given by
\begin{equation}\label{pd:1}
\frac{{\partial J}}{{\partial \theta _m^q}} = 2{{A_q}}\cos \theta _m^q - 2{{B_q}}\sin \theta _m^q - 2{{C_q}}\cos \theta _m^q - 2{{D_q}}\sin \theta _m^q,
\end{equation}
where the expressions of ${A_q}$, ${B_q}$, ${C_q}$, and ${D_q}$ are shown as
\begin{subequations}
\begin{align}
  &\begin{gathered}
  {A_q} = \hfill \\
  \sum\limits_{k = 1}^S {\sum\limits_{j = 1}^S {\sum\limits_{i = 1,i \ne m}^{{Q_q}} {\left| {L_{k,i,j}^q} \right|\left| {L_{k,m,j}^q} \right|\sin \left( {\xi _{k,i,j}^q + \theta _i^q - \xi _{k,m,j}^q} \right),} } }  \hfill \\
\end{gathered}   \hfill  \\
  &{B_q} = \sum\limits_{k = 1}^S {\sum\limits_{j = 1}^S {\left| {L_{k,m,j}^q} \right|\left| {{I_{k,j}}} \right|\sin \left( {\xi _{k,m,j}^q - {\zeta _{k,j}}} \right)} ,}    \\
  &\begin{gathered}
  {C_q} =  \hfill \\
  \sum\limits_{k = 1}^S {\sum\limits_{j = 1}^S {\sum\limits_{i = 1,i \ne m}^{{Q_q}} {\left| {L_{k,i,j}^q} \right|\left| {L_{k,m,j}^q} \right|\cos \left( {\xi _{k,i,j}^q + \theta _i^q - \xi _{k,m,j}^q} \right),} } }  \hfill \\
\end{gathered}         \\
  &{D_q} = \sum\limits_{k = 1}^S {\sum\limits_{j = 1}^S {\left| {L_{k,m,j}^q} \right|\left| {{I_{k,j}}} \right|\cos \left( {\xi _{k,m,j}^q - {\zeta _{k,j}}} \right)} } .  \\
\end{align}
\end{subequations}

By making the partial derivative $\partial J/\partial \theta _m^q$ equal to 0, we can obtain the following two roots within the range of $\theta _m^q \in \left[ {0,2\pi } \right)$
\begin{subequations}\label{ps:1}
\begin{align}
 & \theta _{m,1}^q = \arctan \left( {\frac{{{{A_q}} + {{B_q}}}}{{{{C_q}} - {{D_q}}}}} \right), \label{t1:1}\\
 & \theta _{m,2}^q = \arctan \left( {\frac{{{{A_q}} + {{B_q}}}}{{{{C_q}} - {{D_q}}}}} \right) + \pi . \label{t1:2}
\end{align}
\end{subequations}

The optimal phase shift $\theta _m^q$ is decided by comparing $\theta _{m,1}^q$ and $\theta _{m,2}^q$ to evaluate the phase resulting in a lower objective function $J$.

\begin{remark}
As inferred from (\ref{ps:1}), the optimal phase shift for each meta-atom can be obtained with the knowledge of the channel, connection topology, and the other phase shifts. Moreover, the derived optimal phase shift satisfies $\partial J/\partial \theta _m^q = 0$. Furthermore, the obtained results in (\ref{ps:1}) can be expanded to other generalized channel fitting problems by replacing the identity matrix ${{\mathbf{I}}_S}$ with other matrices. This suggests that by extending the computation of the derivatives to more general scenarios, an SIM is capable of implementing channel fitting flexibly. Compared to conventional optimization algorithms, the derived closed-form solutions ensure convergence to a local optimum while benefiting from lower computational complexity.
\end{remark}

Furthermore, we need to calculate the partial derivative of the objective function $J$ with respect to $\alpha$, thus the objective function is rewritten as
\begin{equation}
J = \left\| {{\mathbf{H}} - \alpha {{\mathbf{I}}_S}} \right\|_F^2,
\end{equation}
where ${\mathbf{H}} = {\mathbf{QGP}}$ denotes the equivalent channel spanning from the transmit ports to the receive ports.

By letting the partial derivative $\partial J/\partial \alpha $ equal to 0, we obtain the unique root as
\begin{equation}\label{alpha:1}
\alpha  = \frac{{{\text{tr}}\left( {{\mathbf{H}} + {{\mathbf{H}}^H}} \right)}}{{2S}}.
\end{equation}

%

By iterating (\ref{ps:1}) and (\ref{alpha:1}) several times, the objective function $J$ progressively converges. For clarity, Algorithm 1 summarizes the detailed procedures of the proposed AO algorithm.

\begin{algorithm} [h]
        \caption{The Proposed AO Algorithm to solve (\ref{pf:1})}
		\label{alg_approx}
		\begin{algorithmic}[1]
        \REQUIRE
			${{\mathbf{W}}^1},{{\mathbf{W}}^2},{{\mathbf{U}}^1},{{\mathbf{U}}^2},{\mathbf{G}},\alpha $.
        \REPEAT
        \FOR {$q = 1;q \leqslant 4;q +  + $}
        \FOR {$m = 1;m \leqslant {Q_q};p +  + $}
           \STATE          Calculating $\theta _{m,1}^q$ and $\theta _{m,2}^q$ via (\ref{t1:1}) and (\ref{t1:2});
           \STATE          Updating $\theta _m^q$ with the one giving the lowest $J$;
        \ENDFOR
        \ENDFOR
           \STATE          Updating $\alpha$ by applying (\ref{alpha:1});
        \UNTIL The decrement of $J$ is lower than a preset threshold or the number of iterations reaches a preset maximum number;
        \ENSURE
			$\alpha,\theta _m^q,q \in \left\{ {1,2,3,4} \right\},m \in \left\{ {1,2, \cdots ,{Q_q}} \right\}$.
		\end{algorithmic}
\end{algorithm}

\begin{remark}
It is worth mentioning that the closed-form solutions presented in (\ref{ps:1}) and (\ref{alpha:1}) guarantee either a monotonic decrease or a constant value of $J$ following each update of $\theta _m^q$ or $\alpha$, thereby guaranteeing monotonic convergence. Furthermore, the final solutions adhere to the KKT conditions as outlined in (\ref{pf:1}), i.e., $\partial J/\partial \theta _m^q = 0,q \in \left\{ {1,2,3,4} \right\},p \in \left\{ {1,2, \cdots ,{Q_q}} \right\}$ and $\partial J/\partial \alpha  = 0$. This implies that any limit point constitutes a locally optimal solution.
\end{remark}

\section{Algorithm Extension for Multi-Layer SIM}

In order to demonstrate the generality of the proposed approach, we extend the previously derived closed-form solution to the conventional multi-layer SIM.

As expected, the closed-form solution for a multiple-layer SIM maintains a similar structure as for a 2-layer SIM. Note that the derived solution also benefits from a lower computational complexity in contrast to the gradient descent (GD) algorithm \cite{SIM1}, as will be shown in the next section.

According to (\ref{UM:1}), the optimization problem for the conventional multi-layer SIM is formulated as

\begin{subequations}\label{pfml:1}
\begin{align}
\mathop {\min }\limits_{\phi _m^l,\varphi _n^k,\alpha } & J = \left\| {{\mathbf{QGP}} - \alpha {{\mathbf{I}}_S}} \right\|_F^2, \label{MLa}\\
{\text{s}}{\text{.t}}{\text{.}} \;\; \;\; &{\mathbf{P}} = {{\mathbf{\Phi }}^L}{{\mathbf{W}}^L} \cdots {{\mathbf{\Phi }}^2}{{\mathbf{W}}^2}{{\mathbf{\Phi }}^1}{{\mathbf{W}}^1}, \label{MLb}\\
&{\mathbf{Q}} = {{\mathbf{U}}^1}{{\mathbf{\Psi }}^1}{{\mathbf{U}}^2}{{\mathbf{\Psi }}^2} \cdots {{\mathbf{U}}^K}{{\mathbf{\Psi }}^K},\label{MLc}\\
&{{\mathbf{\Phi }}^l} = {\text{diag}}\left( {{{\left[ {\phi _1^l,\phi _2^l, \cdots ,\phi _W^l} \right]}^T}} \right),\label{MLd}\\
&{{\mathbf{\Psi }}^k} = {\text{diag}}\left( {{{\left[ {\varphi _1^k,\varphi _2^k, \cdots ,\varphi _U^k} \right]}^T}} \right),\label{MLe}\\
&\left| {\phi _w^l} \right| = 1,m \in \left\{ {1, \cdots ,W} \right\},l \in \left\{ {1, \cdots L} \right\},\label{MLf}\\
&\left| {\varphi _u^k} \right| = 1,n \in \left\{ {1, \cdots ,U} \right\},k \in \left\{ {1, \cdots K} \right\},\label{MLg}\\
&\alpha  \in {\mathbb{R}^*},\label{MLg}
\end{align}
\end{subequations}
where ${{\mathbf{\Phi }}^l}$ denotes the transmission coefficient matrix imposed by the $l$-th layer of TX-SIM, ${{\mathbf{\Psi }}^k}$ represents the transmission coefficient matrix imposed by the $k$-th layer of RX-SIM, ${\phi _w^l}$ stands for the transmission coefficient imposed by the $m$-th meta-atom on the $l$-th layer of TX-SIM, and ${\varphi _u^k}$ is the transmission coefficient imposed by the $n$-th meta-atom on the $k$-th layer of RX-SIM. As inferred from (\ref{MLb}) and (\ref{MLc}), the total number of layers for TX-SIM and RX-SIM are $L$ and $K$, respectively.

Note that the transmission coefficient matrix between the ($l-1$)-th layer of TX-SIM ($l=1$ indicates the transmit ports) and the $l$-th layer of TX-SIM is represented by ${{\mathbf{W}}^l} \in {\mathbb{C}^{W \times W}}$, whose element in the $n$-th row and $\tilde n$-th column represents the corresponding diffraction coefficient between the $n$-th meta-atom on the $l$-th layer and the $\tilde n$-th meta-atom on the ($l-1$)-th layer, i.e., \cite{w1}
\begin{equation}\label{wnn:1}
W_{n,\tilde n}^l = \frac{{{A_t}\cos \chi _{n,\tilde n}^l}}{{r_{n,\tilde n}^l}}\left( {\frac{1}{{2\pi r_{n,\tilde n}^l}} - j\frac{1}{\lambda }} \right){e^{{{j2\pi r_{n,\tilde n}^l} \mathord{\left/
 {\vphantom {{j2\pi r_{n,\tilde n}^l} \lambda }} \right.
 \kern-\nulldelimiterspace} \lambda }}},
\end{equation}
for $\forall n, \tilde n \in \left\{ {1,2, \cdots W} \right\}$, where ${{A_t}}$ denotes the size of the meta-atoms, ${\chi _{n,\tilde n}^l}$ represents the angle between the normal direction of the ($l-1$)-th layer and the propagation direction of EM waves, $\lambda $ indicates the wavelength, and ${r_{n,\tilde n}^l}$ denotes the corresponding transmission distance. In (\ref{wnn:1}), the coupled model between two adjacent metasurfaces is the near-field forward propagation model, and the diffraction coefficient is determined by Rayleigh-Sommerfeld's diffraction theory \cite{w1}.

Similarly, the element in the $n$-th row and $\tilde n$-th column of the transmission coefficient matrix ${{\mathbf{U}}^k} \in {\mathbb{C}^{U \times U}}$ at RX-SIM side is expressed as
\begin{equation}\label{unn:1}
U_{n,\tilde n}^k = \frac{{{A_t}\cos \chi _{n,\tilde n}^k}}{{r_{n,\tilde n}^l}}\left( {\frac{1}{{2\pi r_{n,\tilde n}^k}} - j\frac{1}{\lambda }} \right){e^{j2\pi r_{n,\tilde n}^k{\text{ /}}\lambda }}.
\end{equation}

To begin with, the phase shifts on the $p$-th layer\footnote{For simplicity, the superscript $p$ is labeled by treating TX-SIM and RX-SIM as a unified entity. Therefore, for the case of $p>L$, we have ${{\mathbf{\Phi }}^p} = {{\mathbf{\Psi }}^{L + K - p + 1}}$.} are optimized, and the objective function of (\ref{MLa}) is given by
\begin{equation}\label{ML:1}
J = \left\| {{{\mathbf{A}}^p}{{\mathbf{\Phi }}^p}{{\mathbf{B}}^p} - {\mathbf{I}}} \right\|_F^2,
\end{equation}
where ${{\mathbf{A}}^p}$ denotes the cascaded channel spanning from the $p$-th layer to the receive ports, expressed by
\begin{equation}
{{\mathbf{A}}^p} = \left\{ {\begin{array}{*{20}{c}}
  {{\mathbf{QG}}{{\mathbf{\Phi }}^L}{{\mathbf{W}}^L} \cdots {{\mathbf{\Phi }}^{p + 1}}{{\mathbf{W}}^{p + 1}} \in {\mathbb{C}^{S \times W}},p \leqslant L,} \\
  {{{\mathbf{U}}^1}{{\mathbf{\Phi }}^{L + K}} \cdots {{\mathbf{\Phi }}^{p - 1}}{{\mathbf{U}}^p} \in {\mathbb{C}^{S \times U}},p > L,}
\end{array}} \right.
\end{equation}
${{\mathbf{B}}^p}$ represents the cascaded channel spanning from the transmit ports to the $p$-th layer of TX-SIM (excluding the meta-atoms on the $l$-th layer), given by
\begin{equation}
{{\mathbf{B}}^p} = \left\{ {\begin{array}{*{20}{c}}
  {{{\mathbf{W}}^p}{{\mathbf{\Phi }}^{p - 1}} \cdots {{\mathbf{\Phi }}^1}{{\mathbf{W}}^1} \in {\mathbb{C}^{W \times S}},p \leqslant L,} \\
  {{{\mathbf{U}}^{p + 1}}{{\mathbf{\Phi }}^{p + 1}} \cdots {{\mathbf{U}}^K}{{\mathbf{\Phi }}^{L + 1}}{\mathbf{GP}} \in {\mathbb{C}^{U \times S}},p > L,}
\end{array}} \right.
\end{equation}
and ${\mathbf{I}} = \alpha {{\mathbf{I}}_S}$ indicates the desired fitting channel matrix.

A closed-form solution for the $n$-th phase shift on the $p$-th layer is given by repeating the derivation process outlined in (\ref{Re:1}) to (\ref{ps:1}) as
\begin{subequations}\label{mlps:1}
\begin{align}
 & \vartheta _{n,1}^p = \arctan \left( {\frac{{{{E_p}} + {{F_p}}}}{{{{G_p}} - {{H_p}}}}} \right), \label{mlt1:1}\\
 & \vartheta _{n,2}^p = \arctan \left( {\frac{{{{E_p}} + {{F_p}}}}{{{{G_p}} - {{H_p}}}}} \right) + \pi . \label{mlt1:2}
\end{align}
\end{subequations}
where
\begin{subequations}
\begin{align}
 & {E_p} = \sum\limits_{l = 1}^S {\sum\limits_{j = 1}^S {\sum\limits_{i = 1,i \ne n}^{{Q_p}} {\left| {Y_{l,i,j}^p} \right|\left| {Y_{l,n,j}^p} \right|\sin \left( {\gamma _{l,i,j}^p + \vartheta _i^p - \gamma _{l,n,j}^p} \right)} } } , \label{E:2}\\
 & {F_p} = \sum\limits_{l = 1}^S {\sum\limits_{j = 1}^S {\left| {Y_{l,n,j}^p} \right|\left| {{I_{l,j}}} \right|\sin \left( {\gamma _{l,n,j}^p - {\zeta _{l,j}}} \right)} ,}      \label{F:2}\\
 & {G_p} = \sum\limits_{l = 1}^S {\sum\limits_{j = 1}^S {\sum\limits_{i = 1,i \ne n}^{{Q_p}} {\left| {Y_{l,i,j}^p} \right|\left| {Y_{l,n,j}^p} \right|\cos \left( {\gamma _{l,i,j}^p + \vartheta _i^p - \gamma _{l,n,j}^p} \right),} } }       \label{G:2}\\
 & {H_p} = \sum\limits_{l = 1}^S {\sum\limits_{j = 1}^S {\left| {Y_{l,n,j}^p} \right|\left| {{I_{l,j}}} \right|\cos \left( {\gamma _{l,n,j}^p - {\zeta _{l,j}}} \right)} } ,  \label{H:2}\\
 & Y_{l,i,j}^p = A_{l,i}^pB_{i,j}^p = \left| {Y_{l,i,j}^p} \right|{e^{j\gamma _{l,i,j}^p}},  \label{Y:2}\\
 &{Q_p} = \left\{ {\begin{array}{*{20}{c}}
  {W,p \leqslant L,} \\
  {U,p > L.}
\end{array}} \right. \label{O:2}
\end{align}
\end{subequations}

By iterating (\ref{mlps:1}) several times, the objective function $J$ progressively converges. For the conventional multi-layer SIM, Algorithm 2 summarizes the detailed procedures of the proposed AO algorithm.

\begin{algorithm} [h]
        \caption{The Proposed AO Algorithm to solve (\ref{pfml:1})}
		\label{alg_approx}
		\begin{algorithmic}[1]
        \REQUIRE
			${{\mathbf{W}}^l},l \in \left\{ {1,2, \cdots ,L} \right\},{\mathbf{G}},{{\mathbf{U}}^k},k \in \left\{ {1,2, \cdots ,K} \right\},\alpha $.
        \REPEAT
        \FOR {$p = 1;p \leqslant L+K;p +  + $}
        \FOR {$n = 1;n \leqslant {Q_p};n +  + $}
           \STATE          Calculating $ \vartheta _{n,1}^p $ and $ \vartheta _{n,2}^p $ via (\ref{mlt1:1}) and (\ref{mlt1:2});
           \STATE          Updating $\vartheta  _{n}^p$ with the one giving the lowest $J$;
        \ENDFOR
        \ENDFOR
           \STATE          Updating $\alpha$ by applying (\ref{alpha:1});
        \UNTIL The decrement of $J$ is lower than a preset threshold or the number of iterations reaches a preset maximum number;
        \ENSURE
			$\alpha,\vartheta  _{n}^p$.
		\end{algorithmic}
\end{algorithm}

\begin{remark}
Note that the closed-form solutions presented in (\ref{alpha:1}) and (\ref{mlps:1}) guarantee either a monotonic decrease or a constant value of $J$ for each update of $\varsigma _{n}^p$ or $\alpha$, thereby guaranteeing monotonic convergence. Furthermore, the final solutions adhere to the KKT conditions as shown in (\ref{pfml:1}), indicating that any limit point constitutes a locally optimal solution.
\end{remark}

\section{Performance Analysis}

In this section, we evaluate the channel capacity of the proposed scheme, where the upper bound of the channel capacity is derived to provide insights. Furthermore, we analyze the computational complexity for the proposed scheme and the conventional multi-layer SIM. It is shown that the derived closed-form solutions benefits from linear complexity with respect to the number of meta-atoms per layer.

\subsection{Capacity Analysis}

In this section, we evaluate the channel capacity of the considered SIM-aided MIMO systems operating in additive white Gaussian noise (AWGN). In an ideal scenario, the perfect precoding and combining are carried out by the TX-SIM and RX-SIM to construct a unit channel matrix. However, in practical scenarios, the SIM channel matrix may deviate from the desired identity matrix due to the imperfect fitting of the desired response by using the SIM. This can be formulated as follows:
\begin{equation}
{\mathbf{H}} = \alpha {{\mathbf{I}}_S} + {\mathbf{E}},
\end{equation}
where ${\mathbf{E}}$ denotes the deviation between the SIM channel and the target channel. As a result, the channel capacity when considering the imperfect channel fitting is given by
\begin{equation}
C  \triangleq  {\log _2}\left| {{{\mathbf{I}}_S} + \frac{{{\alpha ^2}{{\mathbf{I}}_S}{{\mathbf{R}}_s}{\mathbf{I}}_s^H}}{{{\mathbf{E}}{{\mathbf{R}}_s}{{\mathbf{E}}^H} + {N_0}{{\mathbf{I}}_S}}}} \right|,
\end{equation}
where ${{{\mathbf{R}}_s}}$ denotes the covariance matrix of the transmit signal, and ${{N_0}}$ represents the power of the AWGN. In this paper, we consider a Gaussian source with covariance matrix ${{\mathbf{R}}_s} = {P_t}{{\mathbf{I}}_S}$. Therefore, the channel capacity is simplified as
\begin{equation}\label{CC:1}
C = {\log _2}\left| {{{\mathbf{I}}_S} + \frac{{{\alpha ^2}{P_t}{{\mathbf{I}}_S}}}{{{\mathbf{E}}{{\mathbf{E}}^H} + {N_0}{{\mathbf{I}}_S}}}} \right|.
\end{equation}

Since it is hard to analytically determine the distribution of ${\mathbf{E}}$, a closed-form expression for (\ref{CC:1}) is unavailable. However, the purpose of SIM is to realize perfect channel fitting, i.e., ${\mathbf{E}} = {\mathbf{0}}$, so that the channel capacity is upper bounded by
\begin{equation}\label{CC:2}
C \leqslant S{\log _2}\left( {1 + {\alpha ^2}{P_t}/{N_0}} \right).
\end{equation}
Eq. (\ref{CC:2}) reveals that when perfect precoding and combining are achieved by TX-SIM and RX-SIM, $S$ parallel SISO sub-channels are constructed to complete the direct transmission through paired transceiver ports.

\subsection{Computational Complexity Analysis}

Next, we analyze the computational complexity of the proposed AO algorithm in terms of number of real-valued multiplications.

For the proposed 2-layer SIM, the computation of the optimal phase shift $\theta _m^q$ with (\ref{ps:1}) has a complexity equal to $\mathcal{O}\left( {{S^2}{Q_q}} \right)$. Moreover, calculating $\alpha$ with (\ref{alpha:1}) entails a complexity of $\mathcal{O}\left( {{S}} \right)$. Let $I$ denote the number of iterations, the total computational complexity of the proposed AO algorithm is $\mathcal{O}\left( {I\left( {5{S^4}{M^2} + 5{S^4}{N^2} + S} \right)} \right)$. In the next section, the number of iterations $I$ is shown through numerical simulations to be a small number under an empirical setup. {The analysis reveals that Algorithm 1 exhibits linear complexity with respect to ${{M^2}}$ and ${{N^2}}$, even for scenarios of large-scale SIM ($M,N \to \infty $).}

{As the number of antennas $S$ increases in large-scale MIMO systems, the required number of meta-atoms in the 2-layer SIM scales linearly with $S$, ensuring hardware scalability. However, the computational complexity grows approximately as $\mathcal{O}\left( {{S^4}} \right)$, which presents challenges for implementation and motivates the development of more efficient optimization algorithms in future work.}

Similarly, for the conventional multi-layer SIM, the computation of the optimal phase shift $\varsigma _{n}^p$ with (\ref{mlps:1}) has a complexity of $\mathcal{O}\left( {{S^2}{Q_p}} \right)$. {As a result, the overall computational complexity of Algorithm 2 is $\mathcal{O}\left( {I\left( {{S^2}{W^2}L + {S^2}{U^2}L + S} \right)} \right)$, which scales linearly with ${{W^2}}$ and ${{U^2}}$, even in large-scale SIM scenarios ($W,U \to \infty $).}

\section{Simulation Results}

In this section, numerical results are provided to evaluate the performance of the proposed 2-layer SIM, and comparisons between the proposed scheme and the conventional multi-layer SIM are also presented to demonstrate the advantages of integrating meta-fiber into the SIM architecture.

\subsection{Simulation Setups}

\begin{table}[t]
\centering
\caption{Simulation Parameters and Descriptions}
\vspace{+5mm}
\begin{tabular}{|c|c|c|}
\hline
\textbf{Parameter} & \textbf{Value} & \textbf{Description} \\
\hline
\multicolumn{3}{|c|}{\textbf{Simulation Setups}} \\
\hline
${P_t}$& 20 dBm&Transmit signal power\\
${N_0}$&$-$110 dBm&Receive noise power\\
$S$&4&Number of data streams\\
$f_0$ & 28 GHz & Carrier frequency\\
$\lambda$ & 10.7 mm & Corresponding wavelength \\
${\beta _0}$ & ${{\text{(4}}\pi {\text{/}}\lambda {\text{)}}^2}$ & Reference path loss\\
$d$ & 150 m & Distance\\
$\gamma$ & 3.5 & Path loss exponent\\

\hline
\multicolumn{3}{|c|}{\textbf{Two-Layer SIM Parameters}} \\
\hline
$M=N$ & 25  & Numbers of meta-atoms \\
$\left| {{\rho_1}} \right| = \left| {{\rho_2}} \right|$& 1 &  Ideal meta-fiber amplitude \\
$\psi_1 = \psi_2$ & 0 & Ideal meta-fiber phase shift \\
\hline
\multicolumn{3}{|c|}{\textbf{Multi-Layer SIM Parameters}} \\
\hline
$L$ & 7 & Conventional SIM layers \\
$W=U$& 100/49 & Meta-atoms per layer \\
${r_{e,t}}$& 5.35 mm & Adjacent meta-atom spacing\\
$d$ & 5.35 mm & Adjacent layer spacing \\
\hline
\multicolumn{3}{|c|}{\textbf{AO Algorithm Initialization}} \\
\hline
${{\mathbf{\Phi }}^l},l \in \mathcal{L}$ & ${{\mathbf{I}}_N}$ & Identity phase shift matrices \\
$\alpha $ & 1 & Initial channel gain \\
$I_{\max}$ & 20 & Maximum number of iterations \\
\hline
\end{tabular}
\label{tab:sim_parameters}
\end{table}

{Simulation parameters are listed in Table \ref{tab:sim_parameters}.} The signal power at the transmitter is set to ${P_t} = 20$ dBm, and the noise power at the receiver is set to ${N_0} =  - 110$ dBm. Moreover, the number of data streams is set to $S = 4$, and the quadrature phase shift keying (QPSK) modulation scheme is considered. Furthermore, the SIM-aided MIMO system operates at the frequency of ${f_0} = 28$ GHz with a wavelength of $\lambda  = 10.7$ mm \cite{SIM2}. The path loss of the MIMO channel, the reference path loss, the distance between the source and destination, and the path loss exponent are set to ${\beta _0}={{\text{(4}}\pi {\text{/}}\lambda {\text{)}}^2}$, $d = 150$ m, and $\gamma = 3.5$, respectively.

For the proposed 2-layer SIM, the numbers of meta-atoms in each sub-area are set to $M=N=25$ unless otherwise specified, which means that the numbers of meta-atoms on the output layer of TX-SIM and the input layer of RX-SIM are $MS=NS=100$. Moreover, the transmission coefficient matrices are expressed as in (\ref{TCM:1}), (\ref{TCM:2}), (\ref{RCM:1}), and (\ref{RCM:2}). For simplicity, the meta-fiber is considered to have no effect on the transmission of the signal, i.e., the ideal case with $\left| {{\rho _1}} \right| = \left| {{\rho _2}} \right| = 1$ and ${\psi _1} = {\psi _2} = 0$ is considered.

For the conventional multi-layer SIM, a 7-layer structure is adopted, and the distance between adjacent layers is set to $\lambda /2$. The number of meta-atoms on each layer is set to $W = U =$ 100 or 49, where the former indicates the same number of meta-atoms on the output layer and the latter means roughly the same (or slightly more) total number of meta-atoms compared to the proposed 2-layer SIM. Accordingly, the transmission coefficients between the adjacent metasurface layers in the SIM are given by (\ref{wnn:1}) and (\ref{unn:1}). Furthermore, a half-wavelength spacing ${r_{e,t}} = \lambda /2$ between adjacent meta-atoms is assumed \cite{SIM1}.

For the proposed AO algorithm, the initial phase shift matrices are set to the identity matrices, i.e., ${{\mathbf{\Phi }}^l} = {{\mathbf{I}}_N},l \in \mathcal{L}$, the initial channel gain $\alpha $ is set to 1, and the maximum allowable number of iterations is set to 20.

For the sake of a fair comparison, a couple of different performance metrics are adopted. Specifically, the normalized mean square error (NMSE) between the SIM channel and the desired identity matrices is defined as
\begin{equation}
\Delta  = \mathbb{E}\left( {\frac{{{{\left\| {{\mathbf{QGP}} - \alpha {{\mathbf{I}}_S}} \right\|}^2}}}{{{{\left\| {\alpha {{\mathbf{I}}_S}} \right\|}^2}}}} \right),
\end{equation}
while the second metric is the channel capacity defined by (\ref{CC:1}).

\subsection{Validation of the Proposed Algorithm}

\begin{figure}[t]
\centering
\includegraphics[width=3.5in]{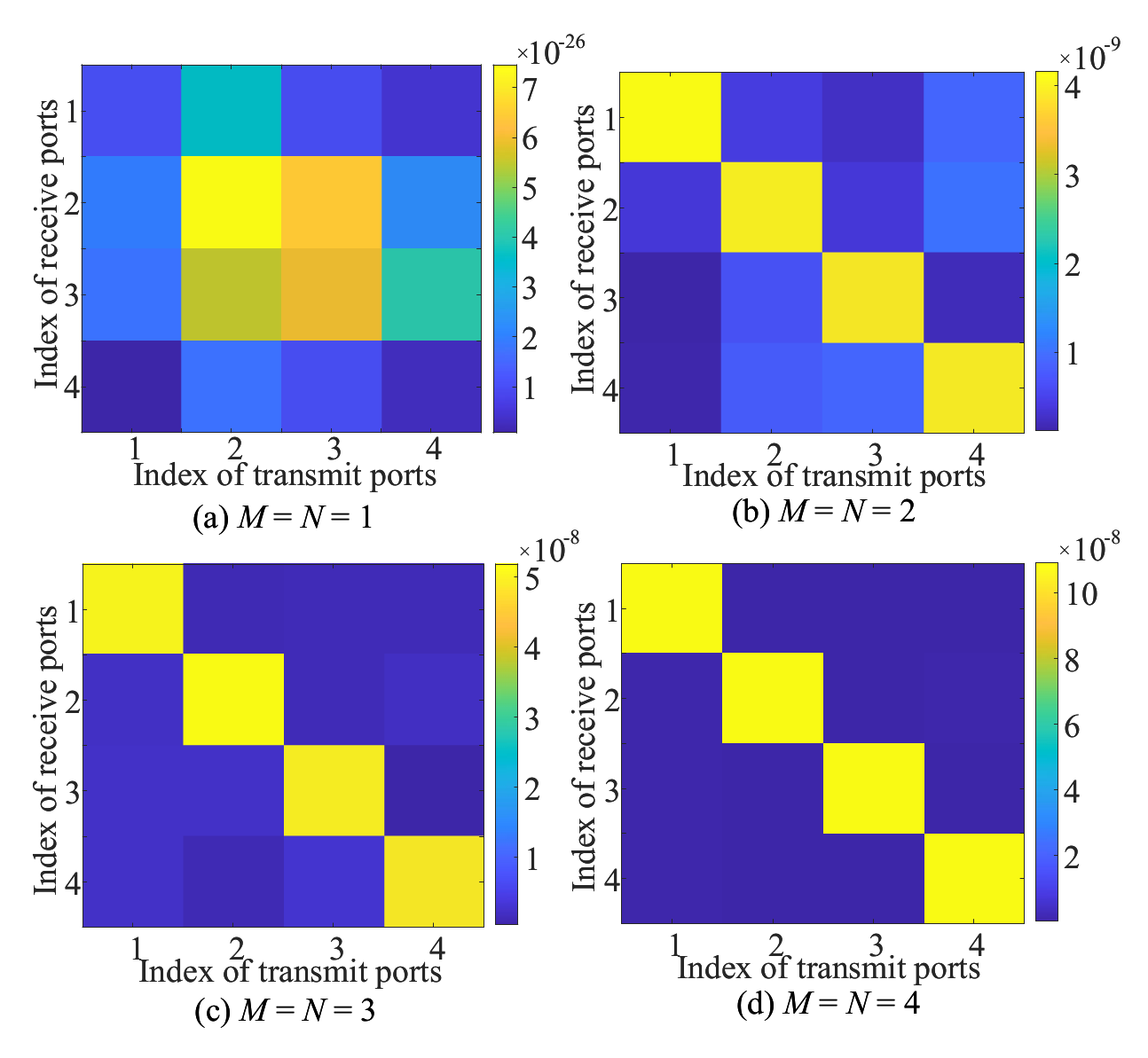}
\vspace{-3mm}
\caption{Visualization of the equivalent channel ${\mathbf{H}} = {\mathbf{QGP}}$.}
\label{fig_3}
\vspace{-0em}
\end{figure}
\begin{figure}[t]
\centering
\includegraphics[width=3.5in,height=2.7in]{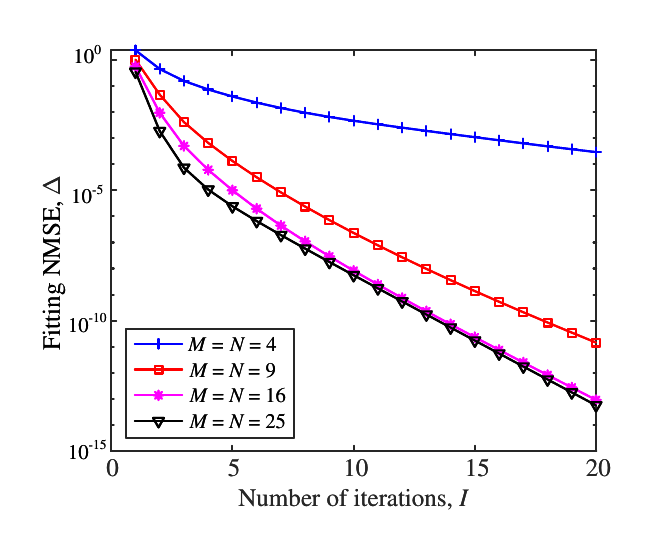}
\vspace{-3mm}
\caption{Convergence curves of the proposed AO algorithms.}
\label{fig_4}
\vspace{-0em}
\end{figure}

In Fig. \ref{fig_3}, the end-to-end channel matrix ${\mathbf{H}} = {\mathbf{QGP}}$ is visualized under different numbers of meta-atoms. As inferred from Fig. \ref{fig_3}, for a small number of meta-atoms, such as $M=N=1$, TX-SIM and RX-SIM are not capable to construct a diagonal channel matrix. This implies that each data stream suffers from severe interference by the others. With an increased number of meta-atoms, i.e., $M=N=4$, the equivalent identity channel is generated. Therefore, each data stream is directly available through the transmit and receive ports, reducing complexity for both the transmitter and receiver.

Fig. \ref{fig_4} evaluates the convergence performance of the proposed AO algorithm. As the number of iterations increases, the NMSE monotonically decreases due to the efficient closed-form solutions derived in (\ref{ps:1}) and (\ref{alpha:1}). After 20 iterations, an NMSE lower than ${10^{ - 10}}$ is achieved assuming $M = N = 9$. Furthermore, increasing the number of meta-atoms results in a lower NMSE.

\emph{In a nutshell, Figs. \ref{fig_3} and \ref{fig_4} verify that, with the proposed algorithm, TX-SIM and RX-SIM construct several parallel sub-channels by appropriately adjusting the phase shifts of the meta-atoms.}

\subsection{Impact of System Parameters}

\begin{figure}[t]
\centering
\includegraphics[width=3.5in,height=2.7in]{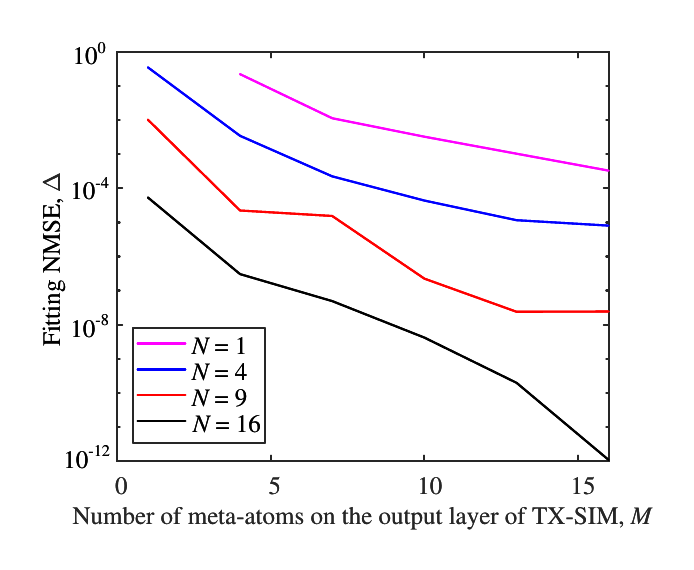}
\vspace{-3mm}
\caption{Influence of the number of meta-atoms on the NMSE.}
\label{fig_5}
\vspace{-0em}
\end{figure}

\begin{figure}[t]
\centering
\includegraphics[width=3.5in,height=2.7in]{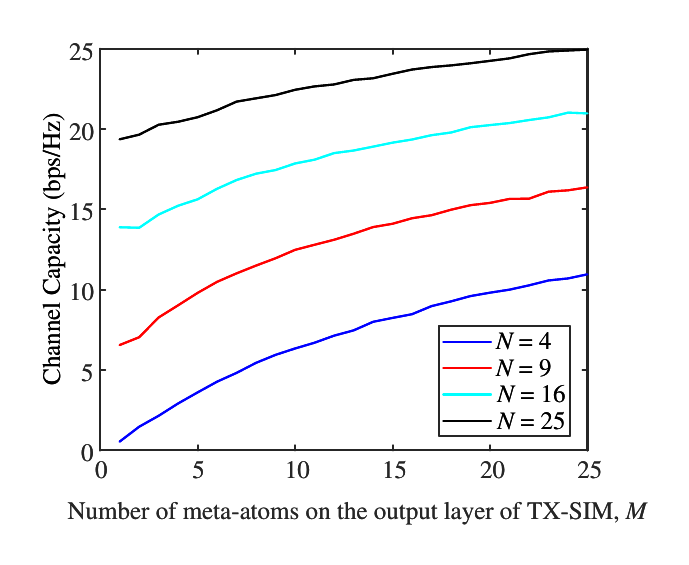}
\vspace{-3mm}
\caption{Influence of the number of meta-atoms on the channel capacity.}
\label{fig_6}
\vspace{-0em}
\end{figure}

Fig. \ref{fig_5} portrays the NMSE between the SIM channel and the desired identity matrices versus different numbers of meta-atoms. As observed from Fig. \ref{fig_5}, when the number of meta-atoms on RX-SIM $N$ is fixed, an increased number of meta-atoms on TX-SIM $M$ contributes to a lower NMSE. Similarly, when the number of meta-atoms on TX-SIM $M$ is fixed, an increased number of meta-atoms on RX-SIM $N$ also contributes to a lower NMSE. In Fig. \ref{fig_6}, the channel capacity versus the number of meta-atoms is illustrated. It can be seen from Fig. \ref{fig_6} that the channel capacity increases with either a larger $M$ or $N$ due to the increased DoF. In general, the more meta-atoms are deployed on TX-SIM and RX-SIM, the lower the NMSE and the higher the channel capacity.

\begin{figure}[t]
\centering
\includegraphics[width=3.5in,height=2.7in]{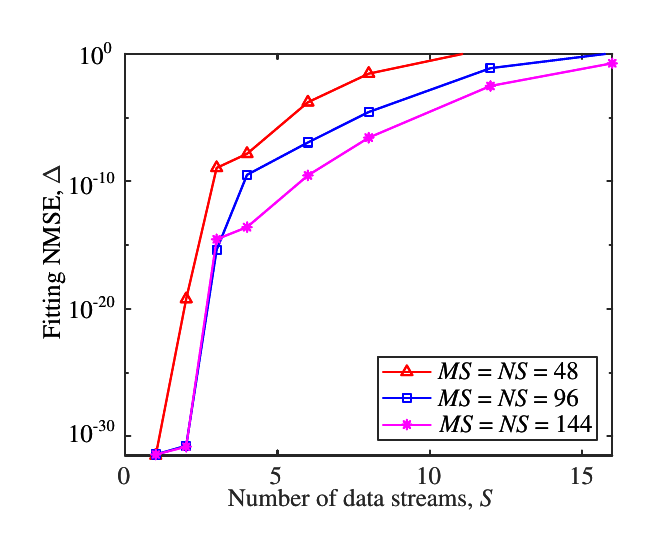}
\vspace{-3mm}
\caption{Impact of the number of data streams on the NMSE.}
\label{fig_7}
\vspace{-0em}
\end{figure}

\begin{figure}[t]
\centering
\includegraphics[width=3.5in,height=2.7in]{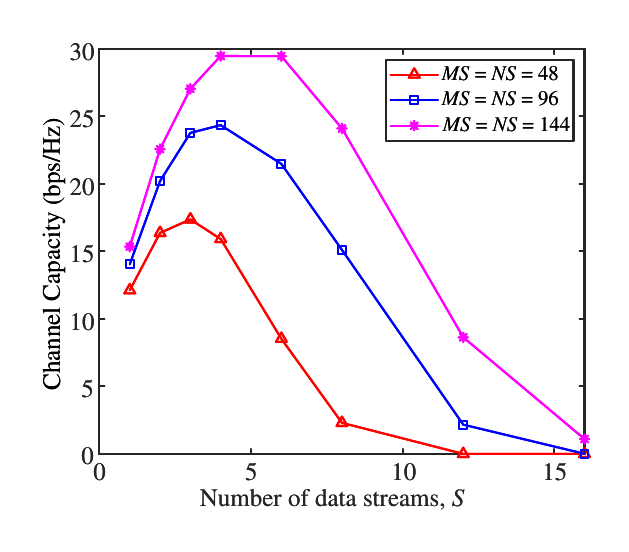}
\vspace{-3mm}
\caption{Impact of the number of data streams on the channel capacity.}
\label{fig_8}
\vspace{-0em}
\end{figure}

In Fig. \ref{fig_7}, the NMSE versus the number of data streams is depicted. As the number of data streams increases, the NMSE also increases. This means that a larger number of data streams $S$ makes fitting the target channel more difficult. Fig. \ref{fig_8} shows the channel capacity versus the number of data streams. As the number of data streams $S$ increases from 1 to 3, the channel capacity increases. However, as $S$ increases from 6 to 16, a large NMSE results in a sharp drop in channel capacity. In summary, an increased number of data streams leads to poorer fitting NMSE. Therefore, maximizing the channel capacity requires a trade-off between channel fitting accuracy and the number of data streams.

\begin{figure}[t]
\centering
\includegraphics[width=3.5in,height=2.7in]{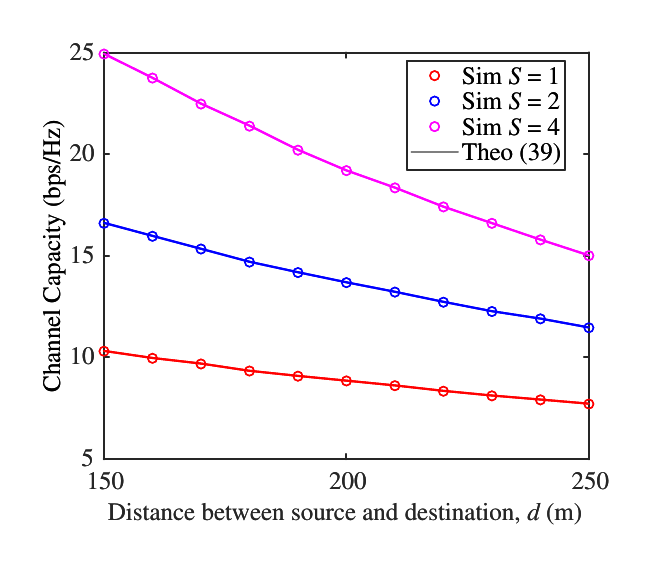}
\vspace{-3mm}
\caption{Influence of distance on the channel capacity.}
\label{fig_9}
\vspace{-0em}
\end{figure}

\begin{figure}[t]
\centering
\includegraphics[width=3.5in,height=2.7in]{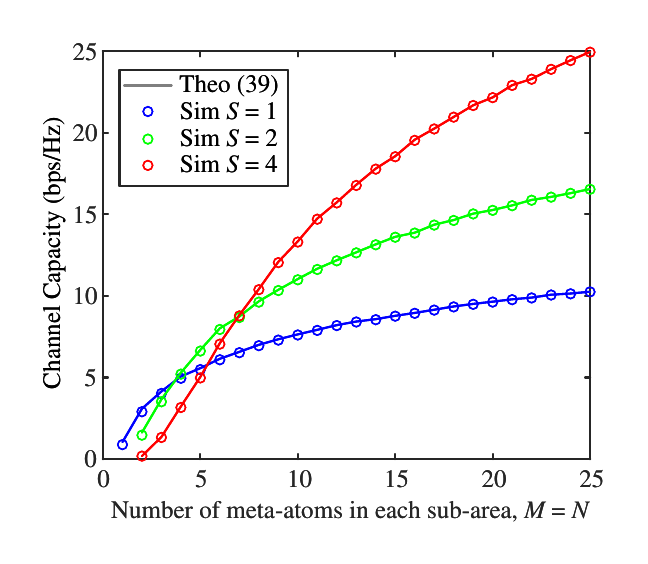}
\vspace{-3mm}
\caption{Influence of the number of meta-atoms on the channel capacity.}
\label{fig_10}
\vspace{-0em}
\end{figure}

Fig. \ref{fig_9} reveals the relationship between the channel capacity and the distance between the source and destination. As expected, as the distance increases, the channel capacity decreases due to increased path loss. Moreover, the theoretical upper bounds in (\ref{CC:2}) tightly approach the simulation results, verifying the correctness of the derivation.

In Fig. \ref{fig_10}, the channel capacity versus the number of meta-atoms in each sub-area is displayed. As observed, when the number of data streams $S$ is fixed, a larger number of meta-atoms $M=N$ contributes to an enhanced channel capacity. Note that for a very small number of meta-atoms $M=N$, a larger number of data streams $S$ may experience lower channel capacity due to the difficulty in fitting the target channel. Furthermore, the tight theoretical bounds and actual simulation results demonstrate the correctness of (\ref{CC:2}).

\emph{In summary, enhanced channel capacity and reduced fitting NMSE are achieved through an increased number of meta-atoms, shorter transceiver distances, and a moderate number of data streams.}

\subsection{Performance Comparison}

\begin{figure}[t]
\centering
\includegraphics[width=3.5in,height=2.7in]{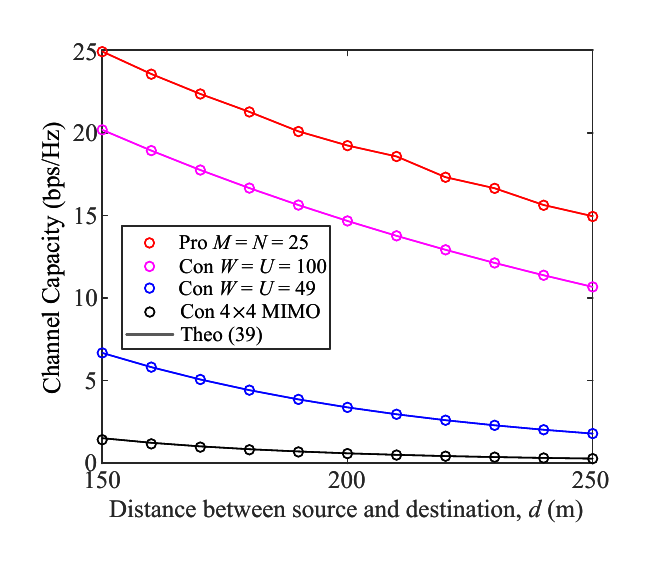}
\vspace{-3mm}
\caption{Comparison of channel capacity versus distance among the proposed 2-layer SIM, conventional 7-layer SIM, and conventional MIMO systems.}
\label{fig_11}
\vspace{-0em}
\end{figure}

\begin{figure}[t]
\centering
\includegraphics[width=3.5in,height=2.7in]{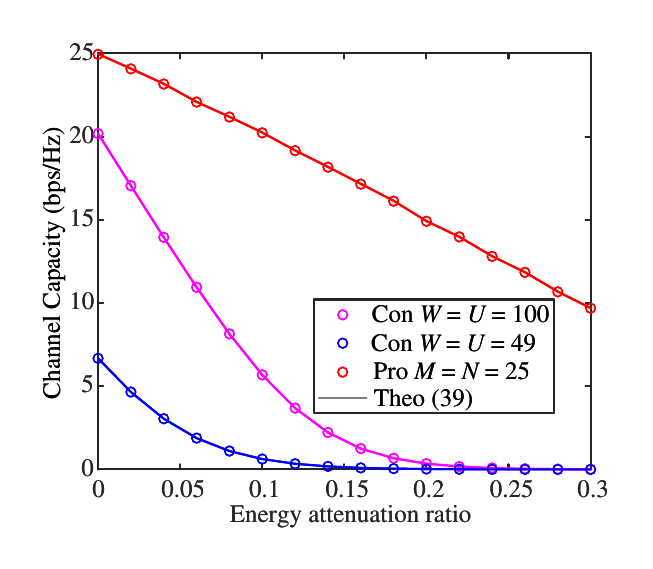}
\vspace{-3mm}
\caption{Comparison of channel capacity versus the energy attenuation ratio between the proposed 2-layer SIM and conventional 7-layer SIM.}
\label{fig_12}
\vspace{-0em}
\end{figure}

\begin{figure}[t]
\centering
\includegraphics[width=3.5in,height=2.7in]{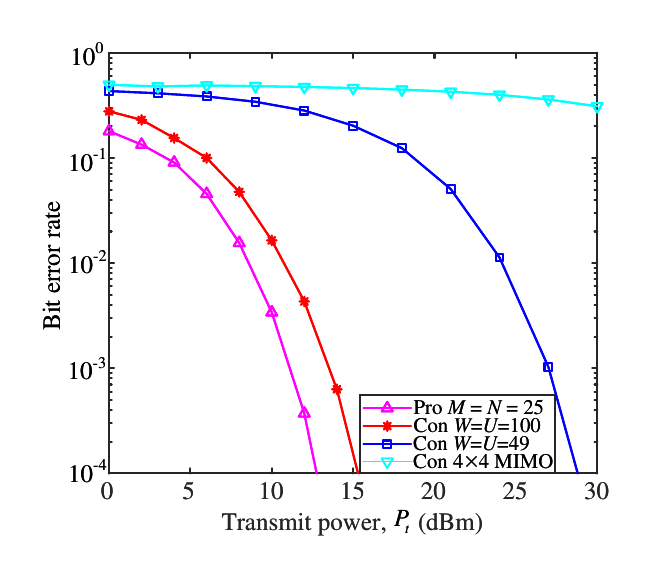}
\vspace{-3mm}
\caption{Comparison of BER among the proposed 2-layer SIM, conventional 7-layer SIM, and conventional MIMO systems.}
\label{fig_BER}
\vspace{-0em}
\end{figure}

{Fig. \ref{fig_11} compares the channel capacity of the proposed 2-layer SIM to that of the conventional 7-layer SIM as well as to its conventional MIMO counterpart.} For the proposed 2-layer SIM scheme, $M=N=25$ implies that the numbers of meta-atoms on the output layer of TX-SIM and input layer of RX-SIM are both $MS=NS=100$. For the conventional 7-layer SIM, $W=U=100$ denotes the equal number of meta-atoms on each layer. Therefore, considering the SIM as a whole, the proposed 2-layer SIM with $M=N=25$ provides the same DoF as the conventional 7-layer SIM with $W=U=100$. {For conventional MIMO systems, we adopt the zero-forcing precoding scheme ${\mathbf{P}} = \sqrt {{P_t}/{\text{Tr}}\left[ {{{\left( {{\mathbf{G}}{{\mathbf{G}}^H}} \right)}^{ - 1}}} \right]} {{\mathbf{G}}^H}{\left( {{\mathbf{G}}{{\mathbf{G}}^H}} \right)^{ - 1}}$ in \cite{ZF1}, which achieves the same channel diagonalization effect as in (\ref{UM:1}). All channels are assumed to be Rayleigh fading along with the pass loss model in (\ref{Gchannel:1}).} As shown in Fig. \ref{fig_11}, the proposed 2-layer SIM benefits from higher channel capacity compared to the conventional 7-layer SIM. The main reason is that the introduction of the meta-fibers avoids the power loss associated with airspace propagation. Although the total number of meta-atoms for the conventional 7-layer SIM with $W=U=49$ is slightly larger than that of the proposed 2-layer SIM, it suffers from reduced channel capacity due to the decreased number of meta-atoms per layer. {The channel capacity of conventional MIMO systems is lower than that of SIM-assisted solutions, as an SIM can dynamically adjust the phases and can leverage inter-atomic diversity. In summary, the proposed 2-layer SIM outperforms both the conventional multi-layer SIM and MIMO counterparts.}

In Fig. \ref{fig_12}, the effect of the energy attenuation ratio on the channel capacity is quantified. The energy attenuation ratio is defined as the signal reduction by each layer of the SIM. In ideal scenarios, ignoring the path loss introduced by Rayleigh-Sommerfield's diffraction, no energy loss is considered when the signal propagate through the layers. However, in actual scenarios, after passing through each layer, some energy is inevitably lost. For example, with an energy attenuation ratio of 0.1, the signal energy after passing through each SIM layer is 90$\%$ of the input. As observed from Fig. \ref{fig_12}, the proposed 2-layer SIM benefits from slower channel capacity attenuation. Specifically, when the energy attenuation ratio is close to 0.25, the conventional 7-layer SIM cannot provide a high channel capacity due to the severe signal attenuation. By contrast, the proposed 2-layer SIM still delivers a channel capacity over 10 bps/Hz. The reduction in the number of layers allows the proposed 2-layer SIM to tolerate greater energy attenuation caused by non-ideal hardware.

{Fig. \ref{fig_BER} compares the bit error rate (BER) performance among the proposed 2-layer SIM, conventional 7-layer SIM, and conventional MIMO systems.} As shown in Fig. \ref{fig_BER}, the proposed 2-layer SIM with $M=N=25$ outperforms the conventional 7-layer SIM with $W=U=100$ under the same number of meta-atoms on the output layer of TX-SIM and the input layer of RX-SIM. As expected, the conventional 7-layer SIM with $W=U=49$ suffers from a higher BER due to the reduction in DoF. {Furthermore, the BER of conventional MIMO systems is higher than that of a 7-layer SIM with $W=U=49$ due to the ability of SIM to dynamically adjust the phases and exploit inter-atomic diversity. Overall, the proposed 2-layer SIM achieves superior BER performance compared to both conventional multi-layer SIM and MIMO systems.}

\emph{In a nutshell, the proposed 2-layer SIM outperforms the conventional 7-layer SIM in terms of channel capacity, BER, and tolerance to energy attenuation. Essentially, the introduction of meta-fibers reduces the power loss associated with signal propagation and provides an effective NMSE fitting with a reduced number of layers.}

\section{Conclusion}
In this paper, we have proposed incorporating meta-fibers into the SIM structure to reduce the number of layers and improve the power efficiency. Specifically, we proposed a 2-layer SIM structure and formulated a channel fitting problem by optimizing the phase shifts. In order to reduce the computational complexity, we designed an efficient AO algorithm to iteratively compute the phase shifts relying on the derived closed-form solutions. Additionally, we provided upper bounds of the channel capacity and quantified the computational complexity. Finally, numerical results demonstrated that the proposed 2-layer SIM achieves over 25$\%$ improvement in channel capacity while reducing the total number of meta-atoms by 59$\%$ compared to the conventional 7-layer SIM. More importantly, this paper puts forth the concept of using meta-fibers to implement SIM, paving the way for the design of novel connection topologies in the future. {Specifically, such novel topologies refer to flexible inter-layer connection designs enabled by meta-fibers, which can facilitate more efficient signal routing across multi-layer SIM structures, potentially reducing the number of required meta-fibers and meta-atoms while enhancing the overall signal processing efficiency. Several important aspects should be carefully considered when integrating meta-fibers into future SIM systems: (1) The number of distinguishable meta-fiber types is inherently finite. In theory, increasing the number of meta-fiber types could enhance the DoF of signals, but this results in exponentially increased fabrication complexity and cost. (2) The transmission coefficient associated with a given meta-fiber remains static over time. If dynamic reconfiguration were feasible, a meta-fiber could offer much better performance. However, such tunable materials remain technologically challenging and are not yet mature for practical implementation. (3) The number of required meta-fibers should scale linearly with the number of meta-atoms to maintain hardware efficiency. If the growth is exponential, it becomes difficult to extend the architecture to support large-scale scenarios, such as in massive MIMO systems, due to the resulting prohibitive spatial and computational overhead.}

\setcounter{subsubsection}{0}

\ifCLASSOPTIONcaptionsoff
  \newpage
\fi

\vspace{-33pt}
\begin{IEEEbiography}[{\includegraphics[width=1in,height=1.25in,clip,keepaspectratio]{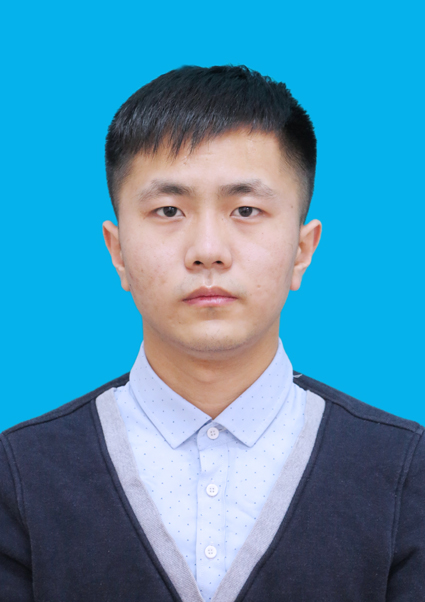}}]{Hong Niu}
received the B.E. and Ph.D. degrees in Information and Communication Engineering from the University of Electronic Science and Technology of China, Chengdu, China, in 2018 and 2024, respectively. He has been a Research Fellow with the School of Electrical and Electronic Engineering, Nanyang Technological University, Singapore. He has published over 10 journals and has participated in several projects. His current research interests include physical-layer security, reconfigurable intelligent surfaces, stacked intelligent metasurfaces, localization, and quantum computing.
\end{IEEEbiography}

\vspace{-33pt}
\begin{IEEEbiography}[{\includegraphics[width=1in,height=1.25in,clip,keepaspectratio]{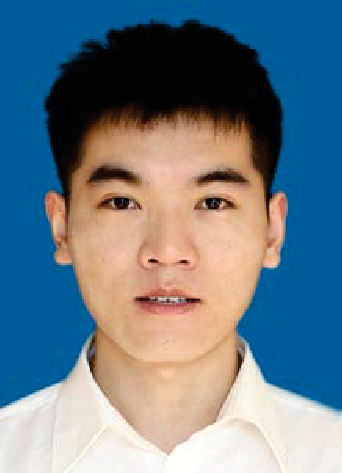}}]{Jiancheng An}
received the B.S. degree in Electronics and Information Engineering and the Ph.D. degree in Information and Communication Engineering from the University of Electronic Science and Technology of China (UESTC), Chengdu, China, in 2016 and 2021, respectively. From 2019 to 2020, he was a Visiting Scholar with the Next-Generation Wireless Group, University of Southampton, U.K. From 2021 to 2023, he was with the Engineering Product Development (EPD) Pillar, Singapore University of Technology and Design. He is currently a Research Fellow with the School of Electrical and Electronics Engineering, Nanyang Technological University (NTU), Singapore. His research interests include stacked intelligent metasurfaces (SIM), wave-based computing, and near-field communications.
\end{IEEEbiography}

\vspace{-33pt}
\begin{IEEEbiography}[{\includegraphics[width=1in,height=1.25in,clip,keepaspectratio]{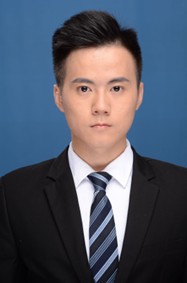}}]{Tuo Wu}
received the B.Eng. degree in telecommunication engineering from South China Normal University, Guangzhou, China, in 2017, and the M.S. degree in wireless radio physics from Sun Yat-Sen University, Guangzhou, China, in 2021. He received the Ph.D. degree from the School of Electronic Engineering and Computer Science, Queen Mary University of London, U.K., in 2024.  He is currently a Postdoctoral Researcher with the School of Electrical and Electronic Engineering, Nanyang Technological University, Singapore. His research interests include Reconfigurable Intelligent Surface, wireless localization, Fluid Antenna Systems, and near-field communications.
\end{IEEEbiography}

\vspace{-33pt}
\begin{IEEEbiography}[{\includegraphics[width=1in,height=1.25in,clip,keepaspectratio]{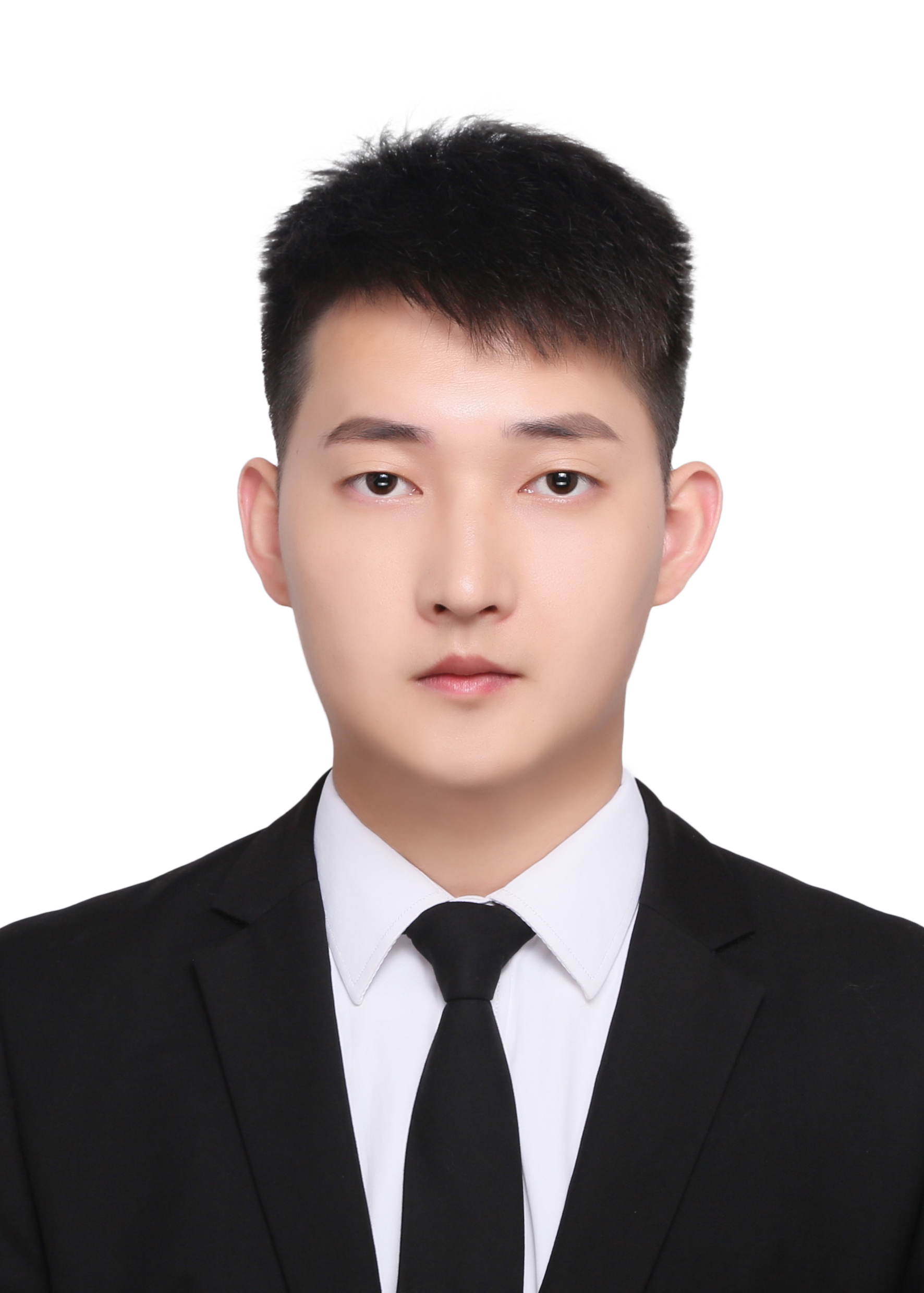}}]{Jiangong Chen}
received the M.S. degree from the University of Electronic Science and Technology of China (UESTC) in 2022, where he is currently pursuing the Ph.D. degree with the National Key Laboratory of Wireless Communications, UESTC. From 2024 to 2025, he was a Visiting Scholar with the Department of Electronic and Electrical Engineering, University College London, London, UK. His current research interests include MIMO systems, reconfigurable intelligent surface, integrated sensing and communication, and physical layer security.
\end{IEEEbiography}

\vspace{-33pt}
\begin{IEEEbiography}[{\includegraphics[width=1in,height=1.25in,clip,keepaspectratio]{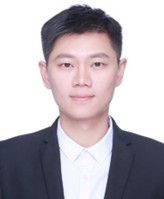}}]{Yufei Zhao}
(Member, IEEE) received the B.Eng. degree (Outstanding Graduates) in Electronic Information Engineering from Harbin Institute of Technology, China, in 2014, and the Ph.D. degree in Aeronautical and Astronautical Science and Technology from Tsinghua University, China, in 2020. From 2020 to 2021, he worked as a Senior Engineer at Huawei Technologies Company Ltd., China. From 2021, he worked as a Research Fellow at the School of Electrical and Electronic Engineering, Nanyang Technological University (NTU), Singapore. His current research interests include wireless communication engineering, electromagnetic science, reconfigurable meta-surface, and orbital angular momentum (OAM) technologies. Dr. Zhao has published more than 50 journal and conference papers. He has 2 filed patents and 5 granted patents. He served as the session Chair for the 2024 IEEE Vehicular Technology Conference (VTC2024-Spring) and the 2024 Asia-Pacific Microwave Conference (APMC 2024). He was also the recipient of the Young Scientist Award of SWA in both 2022 and 2024, and the Best Paper Award of IoTCIT in 2025.
\end{IEEEbiography}

\vspace{-33pt}
\begin{IEEEbiography}[{\includegraphics[width=1in,height=1.25in,clip,keepaspectratio]{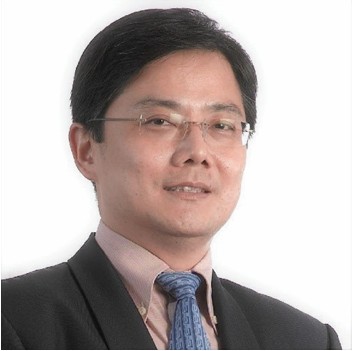}}]{Yongliang Guan}
obtained his PhD degree from the Imperial College London, UK, and Bachelor of Engineering degree with first class honours from
the National University of Singapore. He is now an Associate Vice President of the Nanyang Technological University (NTU), Singapore, and a Professor of Communication Engineering at the School of Electrical and Electronic Engineering in NTU, where
he founded and is leading the Continental-NTU Corporate Research Lab, and also led the successful deployment of the campus-wide NTU-NXP V2X Test Bed in NTU. His research interests broadly include coding and signal processing for communication systems and data storage systems. He has published an invited monograph, 2 books and more than 540 journal and conference papers. He has secured over S\$70 million of external research funding. He has more than 40 filed patents and 13 granted patents (some of which were licensed to NXP, Continental). He is an Editor for the IEEE Transactions on Vehicular Technology and a Distinguished Speaker of the IEEE Vehicular Technology Society.
\end{IEEEbiography}

\vspace{-33pt}
\begin{IEEEbiography}[{\includegraphics[width=1in,height=1.25in,clip,keepaspectratio]{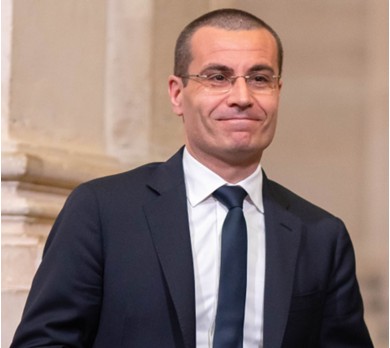}}]{Marco Di Renzo}
(Fellow, IEEE) received the Laurea (cum laude) and Ph.D. degrees in electrical engineering from the University of L'Aquila, Italy, in 2003 and 2007, respectively, and the Habilitation \`a Diriger des Recherches (Doctor of Science) degree from University Paris-Sud (currently Paris-Saclay University), France, in 2013. Currently, he is a CNRS Research Director (Professor) and the Head of the Intelligent Physical Communications group with the Laboratory of Signals and Systems (L2S) at CNRS \& CentraleSup\'elec, Paris-Saclay University, Paris, France, as well as a Chair Professor in Telecommunications Engineering with the Centre for Telecommunications Research -- Department of Engineering, King's College London, London, United Kingdom. He was a France-Nokia Chair of Excellence in ICT at the University of Oulu (Finland), a Tan Chin Tuan Exchange Fellow in Engineering at Nanyang Technological University (Singapore), a Fulbright Fellow at The City University of New York (USA), a Nokia Foundation Visiting Professor at Aalto University (Finland), and a Royal Academy of Engineering Distinguished Visiting Fellow at Queen's University Belfast (U.K.). He is a Fellow of the IEEE, IET, EURASIP, and AAIA; an Academician of AIIA; an Ordinary Member of the European Academy of Sciences and Arts, an Ordinary Member of the Academia Europaea, and Ordinary Member of the Italian Academy of Technology and Engineering; an Ambassador of the European Association on Antennas and Propagation; and a Highly Cited Researcher. His recent research awards include the Michel Monpetit Prize conferred by the French Academy of Sciences, the IEEE Communications Society Heinrich Hertz Award, and the IEEE Communications Society Marconi Prize Paper Award in Wireless Communications. He served as the Editor-in-Chief of IEEE Communications Letters from 2019 to 2023. His current main roles within the IEEE Communications Society include serving as a Voting Member of the Fellow Evaluation Standing Committee, as the Chair of the Publications Misconduct Ad Hoc Committee, and as the Director of Journals. Also, he is on the Editorial Board of the Proceedings of the IEEE.
\end{IEEEbiography}

\vspace{-33pt}
\begin{IEEEbiography}[{\includegraphics[width=1in,height=1.25in,clip,keepaspectratio]{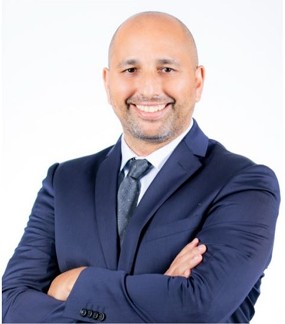}}]{M\'{e}rouane Debbah}
(Fellow, IEEE) is a Professor at Khalifa University of Science and Technology in Abu Dhabi and founding Director of the KU 6G Research Center. He is affiliated as an Adjunct Professor with the Department of the School of Electrical Engineering, Korea University. He is a frequent keynote speaker at international events in the field of telecommunication and AI. His research has been lying at the interface of fundamental mathematics, algorithms, statistics, information and communication sciences with a special focus on random matrix theory and learning algorithms. In the Communication field, he has been at the heart of the development of small cells (4G), Massive MIMO (5G) and Large Intelligent Surfaces (6G) technologies. In the AI field, he is known for his work on Large Language Models, distributed AI systems for networks and semantic communications. He received multiple prestigious distinctions, prizes and best paper awards (more than 40 IEEE best paper awards) for his contributions to both fields and according to research.com is ranked as the best scientist in France in the field of Electronics and Electrical Engineering. He is an IEEE Fellow, a WWRF Fellow, a Eurasip Fellow, an AAIA Fellow, an Institut Louis Bachelier Fellow and a Membre \'{e}m\'{e}rite SEE.
\end{IEEEbiography}

\vspace{-33pt}
\begin{IEEEbiography}[{\includegraphics[width=1in,height=1.25in,clip,keepaspectratio]{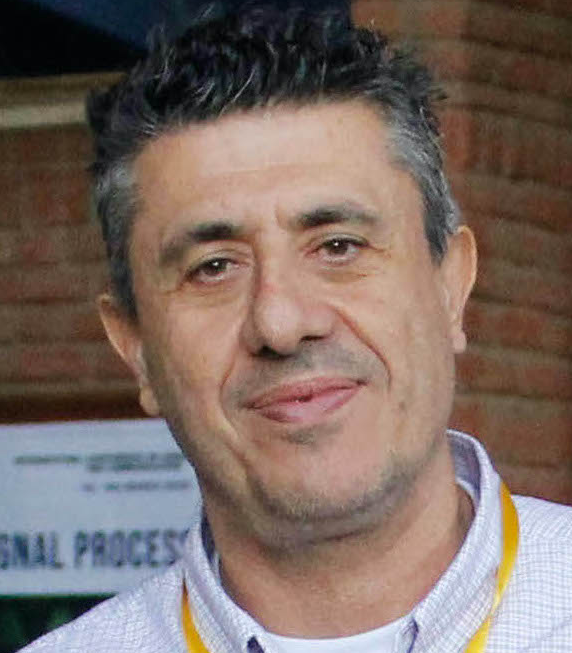}}]{George K. Karagiannidis}
(IEEE Fellow) received the Ph.D. degree in Telecommunications Engineering from Electrical Engineering Department, University of Patras, Greece, in 1998. He is currently a Professor with the Electrical and Computer Engineering Department, Aristotle University of Thessaloniki, Thessaloniki, Greece, and the Head of Wireless Communications and Information Processing (WCIP) Group. His research interests are in the areas of wireless communications systems and networks, signal processing, optical wireless communications, wireless power transfer, and signal processing for biomedical engineering.

Dr. Karagiannidis recently received three prestigious awards: The 2021 IEEE ComSoc RCC Technical Recognition Award, the 2018 IEEE ComSoc SPCE Technical Recognition Award, and the 2022 Humboldt Research Award from Alexander von Humboldt Foundation. He is one of the Highly Cited Authors across all areas of Electrical Engineering, recognized from Clarivate Analytics as the Web-of-Science Highly-Cited Researcher in the ten consecutive years 2015¨C2024. Currently, he is the Editor-in Chief of IEEE Transactions on Communications and in the past was the Editor-in Chief of IEEE Communications Letters.
\end{IEEEbiography}

\vspace{-33pt}
\begin{IEEEbiography}[{\includegraphics[width=1in,height=1.25in,clip,keepaspectratio]{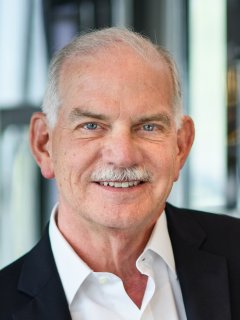}}]{H. Vincent Poor}
(S'72, M'77, SM'82, F'87) received the Ph.D. degree in EECS from Princeton University in 1977.  From 1977 until 1990, he was on the faculty of the University of Illinois at Urbana-Champaign. Since 1990 he has been on the faculty at Princeton, where he is currently the Michael Henry Strater University Professor. During 2006 to 2016, he served as the dean of Princeton's School of Engineering and Applied Science, and he has also held visiting appointments at several other universities, including most recently at Berkeley and Caltech. His research interests are in the areas of information theory, machine learning and network science, and their applications in wireless networks, energy systems and related fields. Among his publications in these areas is the book Machine Learning and Wireless Communications.  (Cambridge University Press, 2022). Dr. Poor is a member of the National Academy of Engineering and the National Academy of Sciences and is a foreign member of the Royal Society and other national and international academies. He received the IEEE Alexander Graham Bell Medal in 2017.
\end{IEEEbiography}

\vspace{-33pt}
\begin{IEEEbiography}[{\includegraphics[width=1in,height=1.25in,clip,keepaspectratio]{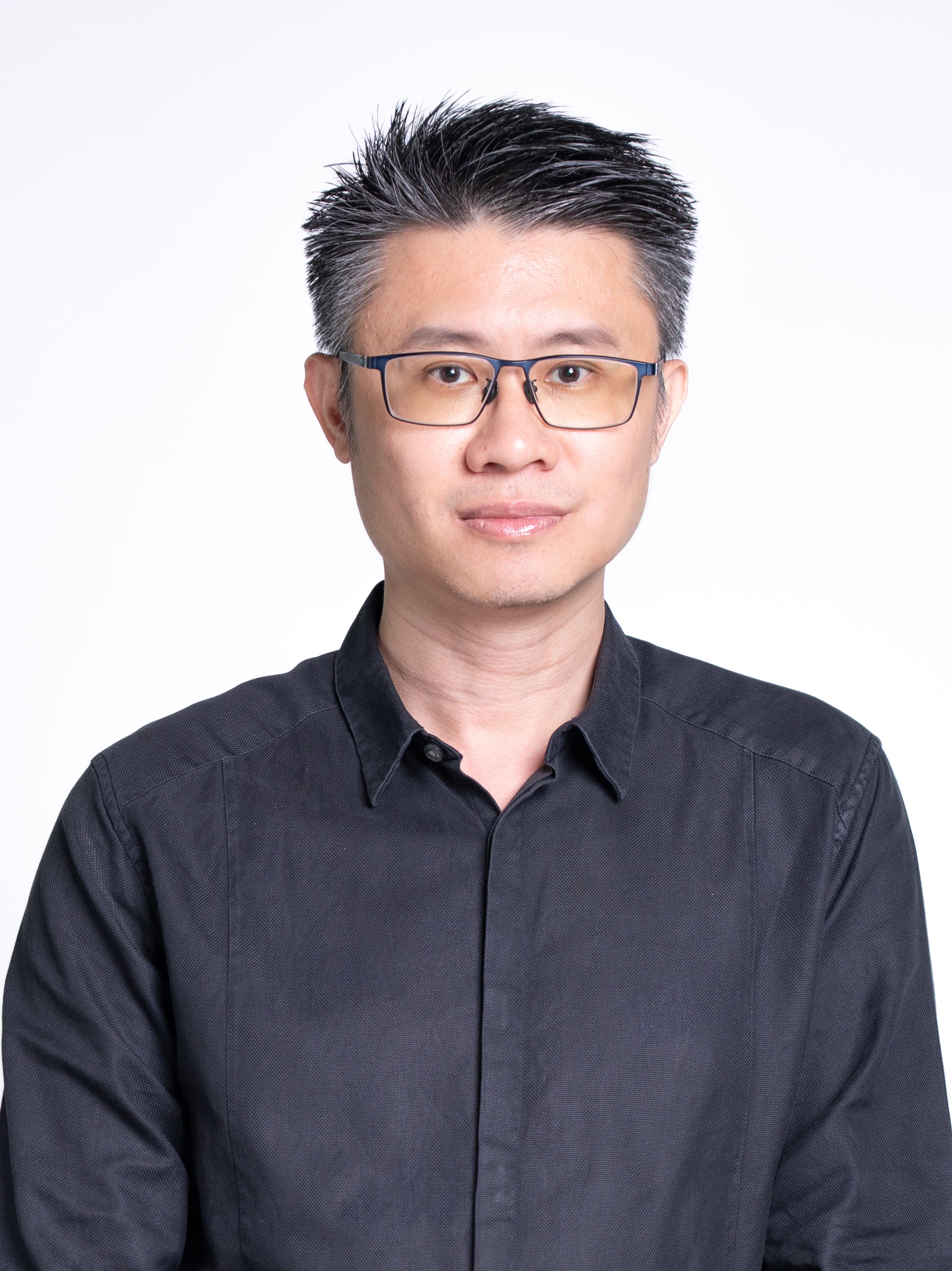}}]{Chau Yuen}
(S02-M06-SM12-F21) received the B.Eng. and Ph.D. degrees from Nanyang Technological University, Singapore, in 2000 and 2004, respectively. He was a Post-Doctoral Fellow with Lucent Technologies Bell Labs, Murray Hill, in 2005. From 2006 to 2010, he was with the Institute for Infocomm Research, Singapore. From 2010 to 2023, he was with the Engineering Product Development Pillar, Singapore University of Technology and Design. Since 2023, he has been with the School of Electrical and Electronic Engineering, Nanyang Technological University, currently he is Provost's Chair in Wireless Communications, Assistant Dean in Graduate College, and Cluster Director for Sustainable Built Environment at ER@IN.

Dr. Yuen received IEEE Communications Society Leonard G. Abraham Prize (2024), IEEE Communications Society Best Tutorial Paper Award (2024), IEEE Communications Society Fred W. Ellersick Prize (2023), IEEE Marconi Prize Paper Award in Wireless Communications (2021), IEEE APB Outstanding Paper Award (2023), and EURASIP Best Paper Award for JOURNAL ON WIRELESS COMMUNICATIONS AND NETWORKING (2021).

Dr. Yuen current serves as an Editor-in-Chief for Springer Nature Computer Science, Editor for IEEE TRANSACTIONS ON VEHICULAR TECHNOLOGY, IEEE TRANSACTIONS ON NEURAL NETWORKS AND LEARNING SYSTEMS, and IEEE TRANSACTIONS ON NETWORK SCIENCE AND ENGINEERING, where he was awarded as IEEE TNSE Excellent Editor Award 2024 and 2022, and Top Associate Editor for TVT from 2009 to 2015. He also served as the guest editor for several special issues, including IEEE JOURNAL ON SELECTED AREAS IN COMMUNICATIONS, IEEE WIRELESS COMMUNICATIONS MAGAZINE, IEEE COMMUNICATIONS MAGAZINE, IEEE VEHICULAR TECHNOLOGY MAGAZINE, IEEE TRANSACTIONS ON COGNITIVE COMMUNICATIONS AND NETWORKING, and ELSEVIER APPLIED ENERGY.

He is listed as Top 2\% Scientists by Stanford University, and also a Highly Cited Researcher by Clarivate Web of Science from 2022. He has 4 US patents and published over 500 research papers at international journals.
\end{IEEEbiography}

\vfill

\end{document}